\def\grs{GRS~$1915$+$105$}
\def\X1550{XTE~J$1550-564$}
\def\J1655{GRO~J$1655-40$}

\def\integral{{\it{INTEGRAL}}}
\def\rxte{{\it{RXTE}}}

\documentclass[]{aastex}
\usepackage{emulateapj5}
\usepackage{amssymb}
\usepackage{natbib}
\usepackage{epsfig}
\usepackage{epsf}
\usepackage{color}
\definecolor{red}{rgb}{0.7,0,0}
\definecolor{blue}{rgb}{0,0,0.7}
\def\correc#1{{#1}}

\shorttitle{An INTEGRAL monitoring of GRS~1915+105: Part1}
\shortauthors{Rodriguez et al.}
\begin{document}
\title{Two years of INTEGRAL monitoring of GRS~1915+105 \\
Part 1: multiwavelength coverage with INTEGRAL, RXTE, and the Ryle radio 
Telescope}

\author{J. Rodriguez\altaffilmark{1}, D.C. Hannikainen\altaffilmark{2}, S.E. Shaw\altaffilmark{3}, G. Pooley\altaffilmark{4}, S. Corbel\altaffilmark{1}, M. Tagger\altaffilmark{5}, I.F. Mirabel\altaffilmark{6}, T. Belloni\altaffilmark{7}, C. Cabanac\altaffilmark{3}, M. Cadolle Bel\altaffilmark{8}, J. Chenevez\altaffilmark{9}, P. Kretschmar\altaffilmark{8}, H.J. Lehto\altaffilmark{10}, A. Paizis\altaffilmark{11}, P. Varni\`ere\altaffilmark{12}, O. Vilhu\altaffilmark{2}
}

\altaffiltext{1}{Laboratoire AIM, CEA/DSM - CNRS - Universit\'e Paris Diderot, DAPNIA/SAp, F-91191 Gif-sur-Yvette, France}
\altaffiltext{2}{Observatory, PO Box 14, FI-00014 University of Helsinki, Finland}
\altaffiltext{3}{School of Physics and Astronomy, University of Southampton, SO17 1BJ, UK}
\altaffiltext{4}{Astrophysics, Cavendish Laboratory, J J Thomson Avenue, Cambridge CB3 0HE, UK}
\altaffiltext{5}{Service d'Astrophysique, (UMR AstroParticules et Cosmologie), CEA Saclay
91191 Gif-sur-Yvette, France}
\altaffiltext{6}{European Southern Observatory, Chile. On leave from CEA-Saclay, France}
\altaffiltext{7}{INAF-Osservatorio Astronomico di Brera, via Bianchi 46, 23807 Merate, Italy}
\altaffiltext{8}{European Space Astronomy Centre (ESAC)
Apartado/P.O. Box 78, Villanueva de la Ca\~nada, E-28691 Madrid, Spain}
\altaffiltext{9}{Danish National Space Center, Technical University of Denmark, Juliane Maries Vej 30, 2100 Copenhagen, Denmark}
\altaffiltext{10}{Tuorla Observatory and  Department of Physics, University of Turku V\"ais\"al\"antie 20, FI-21500 Piikki\"o, Finland}
\altaffiltext{11}{IASF Milano-INAF, Via Bassini 15, 20133 Milano, Italy}
\altaffiltext{12}{LAOG, Universit\'e J. Fourier (UMR5571), Grenoble, France}

\begin{abstract}
We report the results of monitoring 
observations of the Galactic microquasar \grs\ performed 
simultaneously with \integral\ and \rxte\ from 3 up to $\sim$300 keV, and 
the Ryle Telescope at 15 GHz. 
We present the results of the whole \integral\ campaign,  report the 
sources that are detected and their fluxes and identify the classes of 
variability in which \grs\ is found. \correc{The long and continuous \integral\
exposures enable us to see several direct transitions between different classes
of variability.} We focus on the connection between the 
behavior of \grs\ at X-ray energies and at radio wavelengths. \correc{The 
data are studied in a model independent manner through the source light curves, its   
hardness ratio, and color color diagrams}. During a period of 
steady ``hard'' X-ray state (the so-called class $\chi$) we observe a steady radio 
flux. This is interpreted as the signature of the presence of a compact jet.
We then turn to 3 particular observations during which we observe several types 
of soft X-ray dips and spikes cycles, followed by radio flares.
During these observations \grs\ is in the so-called $\nu$, $\lambda$, and 
$\beta$ classes of variability. The observation of ejections 
during class $\lambda$ are the first ever reported. \correc{Our model independent approach 
of the high energy data allows us} to generalize the fact that a (non-major) 
discrete ejection always occurs, in \grs, as a response to an X-ray sequence 
composed of a \correc{spectrally hard} X-ray dip (more pronounced at soft X-rays) 
terminated by an X-ray spike \correc{marking the disappearance of the hard X-ray 
emission above 18~keV.} \correc{This model independent approach also permits us to }
identify the trigger of the ejection as this X-ray spike. In addition, 
a possible correlation between the amplitude of the radio flare and the duration 
of the X-ray 
dip is found in our data. In this case the X-ray dips prior to ejections could be seen as 
the time during which the source accumulates energy and material that is ejected later.
\correc{The fact that these results do not rely on any spectral modelling enhances their 
robustness.}
\end{abstract}
\keywords{accretion, accretion disks --- black hole physics --- stars: individual 
(GRS 1915+105) --- X-rays: binaries --- radio continuum: stars}

\section{Introduction}
\indent Microquasars are the Galactic scaled-down versions of AGNs 
\correc{\citep{m&r98}}. 
In both classes of systems, the copious emission of energy is thought to 
originate from the accretion of matter onto the central black hole (BH), 
which occurs through an accretion disk. Relativistic ejections are 
observed in both classes 
of objects, either through discrete jets, or in a self-absorbed compact jet. 
Apart from morphological similarities, the difference of the mass of the central
object leads to a higher temperature of the inner regions of the 
accretion disk in the Galactic sources, smaller extent of the jets, and, of high 
importance, smaller 
time scales in any of the phenomena associated to either accretion or ejection 
processes \correc{\citep{m&r98}}. As a result, microquasars are excellent laboratories to study the 
accretion-ejection links on time scales from seconds to days. This is done by  coupling 
the variations seen at X-ray energies (mapping the regions closest to the compact object)
to those seen at radio and infrared (IR) wavelengths (representing the emission from the jets).
Such a work has been  initiated  in GRS 1915+105 \citep{pooley97,eikenberry98,mirabel98,fender98}, 
and pursued in a large number of systems since then.\\ 
\indent An extensive review on \grs\ can be found in \citet{fender04}.
To summarize, \grs\ hosts a BH of  14.0$\pm$4.4 M$_{\odot}$ 
\citep{harl04},
it is one of the brightest X-ray sources in the sky and a source of superluminal 
ejections \citep{m&r94}, with true bulk velocity $\geq 0.9$c.
From these superluminal motions an upper limit on the distance to \grs\
  of 11.2 kpc could be derived (on the assumption of intrinsic symmetry 
\correc{in the bipolar jets} \citet{fender99}), 
although a distance as low as  
6 kpc \citep{chapuis04} cannot be excluded. The source is also known to 
show a compact jet during its periods of low and steady level of X-ray 
emission \citep{dhawan}. \\
\indent \grs\ has been
extensively observed with the {\it Rossi X-ray Timing Explorer} ({\it
RXTE}) since 1996.  A rich pattern of variability has emerged from
these data with time scales from years down to 15~ms \citep[e.g.][]{morgan}. 
\citet{belloni00}, analyzing 163 {\it
RXTE} observations from 1996--97, classified all the observations into
12 separate classes (labeled with greek letters) based on count rates and color
characteristics. This scheme has been widely used ever since and is
also applied here. The classes could be interpreted as transitions between
three basic states (A-B-C): A being equivalent to the Soft State, B to the Soft
Intermediate State and C to the Hard Intermediate state in the classification
of \cite{homan06}. These spectral changes are, in most of the classes, interpreted 
as reflecting the rapid disappearance of the inner portions of an accretion disc, 
followed by a slower refilling of the emptied region \citep{belloni97}. Note that other 
possibilities as the disappearance of the corona \citep{chaty98,rod02b,vadawale}, or 
dissipation of magnetic energy \citep[e.g.][]{tagger04}  have also been invoked 
 in some models. \\
\indent Multi-wavelength coverages involving radio, IR and X-ray telescopes have shown a
clear but complex association between soft X-rays and radio emission,
including radio QPOs in the range 20--40 min which were associated with X-ray 
variations on the same time scale \citep{pooley97,eikenberry98,mirabel98,fender98}.  
These oscillations were ascribed to small ejections of material from the
system, and were found to correlate with the disk instability, as
observed in the X-ray band.  This was the first time that the disk-jet
relation could be studied in detail. This kind of cycle
could also reflect some magnetic flood scenario  in
which reconnection events would allow the ejection of blobs of
material \citep{tagger04}.\\
\indent While fine X-ray spectral and temporal analysis will be presented in a companion
paper (Rodriguez et al. 2007 hereafter paper 2), we analyze here , \correc{in a model 
independent way}, the multi-wavelength 
data from our 2 years monitoring campaign focusing on the observations during which correlated
X-ray and radio variabilities are seen. We start with giving the basic properties
of our \integral\ observations, i.e. we describe the campaign and the data reduction 
processes in section 2. In section 3 and 4 we  describe and discuss the four observations 
from which strong radio/X-ray connections are observed. 

\section{Observations and Data Reduction}
\subsection{Journal of the observations}
The journal of the \integral\ observations belonging to our monitoring campaign taken 
from 2004 October to 2005 December is given in Table \ref{tab:journalinteg}, while 
Table \ref{tab:logrxte} reports 
the details of our simultaneous \rxte\ monitoring.  
The daily multiwavelength light curves of \grs\ as seen by \rxte/ASM (1.2-12 keV), the 
Ryle Telescope (RT, 15 GHz) over the period of interest are reported in  Fig. 
\ref{fig:lite} (January 1$^{\rm{st}}$ 2004 is 
MJD 53 005). The days of our \integral\ 
pointings are shown as vertical arrows, the four bigger ones indicate the four 
observations whose accretion ejection properties are discussed in more detail
in this paper. 

\subsection{\integral\ data reduction}
Our monitoring makes use of the \integral\ Soft Gamma-Ray Imager (ISGRI, \citet{lebrun03}) 
-- the low energy detector of the Imager On-Board \integral (IBIS) -- to cover the 18 
to $\sim300$ keV energy range, and the X-ray monitors JEM-X \citep{lund03} to cover the 
3-30 keV energy range. \correc{Both instruments see the sky through a coded mask.} 
We reduced the \integral\ data using the Off Line Scientific 
Analysis ({\tt OSA}) \correc{v. 7.0}. Apart from the different version of the softwares, the 
data from ISGRI and JEM-X were reduced in 
a way similar to that presented in \citet{rod05} for another source present in the 
totally coded field 
of view (TCFOV) of ISGRI, IGR~J19140+0951. Our first step was to produce ISGRI images 
in two energy bands,
20--40 and 40--80 keV, in order to detect the active sources of the FOV which have 
to be taken into account in the (spectral and temporal) extraction processes. 
All sources with a signal to noise ratio greater than 6 in the 20--40 keV ranges 
were considered. The list of sources found in each revolution  is given in 
Table \ref{tab:Listsource}. Apart from a purely technical 
process of obtaining the cleanest data for the main target of our analysis 
\citep[e.g.][for the details of the IBIS data reduction processes]{goldwurm03}, this 
procedure also allows us to survey the brightest sources of 
this field and analyze their data \correc{(e.g. analysis of Aql X-1 during its faint April 
2005 outburst can be found in \citet{rod06})}\\
\indent Since \grs\ is the main target of all our observations,  with an  
hexagonal dithering pattern  it is always in the TCFOV of ISGRI, and 
in the useful part of the FOV of JEM-X. We, therefore, did not include any  restriction 
concerning the off-axis angle in selecting the data.  The second step of our analysis was to 
produce JEM-X and ISGRI light curves with the view to identify the class, and monitor 
the behavior of the source on short time scales. The JEM-X light curves 
were extracted with a time resolution of 1s \correc{between $\sim3$--$13$ keV, and 
further rebinned to 20~s}. The ISGRI light curves were 
extracted \correc{between 18--50~keV with a time resolution of 20~s.
They were further rebinned as they can be quite noisy and hide interesting patterns 
when plotted on such a short time bin. The rebinning depended on the brightness of the source 
at energies above 18~keV and was usually in the range 50-- 200~s.}
\correc{The instrumental and sky background is estimated on the same 
time bins as the source raw count rate, from non-illuminated pixels. For ISGRI an additional
step is to renormalize this background using background maps provided with the calibration tree. 
 For each instrument, the light curves are corrected for background during the extraction process.}
From these background corrected light curves we produced softness ratios (SR) defined as SR= 3--13~keV/18--50~keV. 
\subsection{\rxte\ data reduction}
The \rxte\ data were reduced with the {\tt LHEASOFT} v. 6.1.2.
\correc{All data products were extracted from user's good times intervals (GTI).
GTIs corresponded to times when the satellite elevation was greater than 
10$^\circ$ above the Earth limb, the offset pointing less 
than 0.02$^\circ$, and proportional counter unit \#2 was active.} 
In order to identify the classes, 
we extracted 1s resolution light curves in the 2--60 keV range and 
in the three energy bands defined in \citet{belloni00}, 
from the \correc{Proportional Counter Array (PCA)}.  These bands are  2--5.7 keV 
(channels 0-13, PCA epoch 5), 5.7--14.8 keV 
(channels 14-35), and $>$14.8 keV (channels 36-255). The colors were defined as 
HR1=5.7--14.8/2--5.7 keV and HR2=14.8--60/2--5.7 keV. The shift of gain between the different 
epochs of PCA leads to different absolute values of the count rates, hardness ratios (HR) and 
position in the color-color (CC) diagrams, but the general shape of a given class is easily 
comparable to those of epoch 3 \citep{belloni00}, and therefore allowed us to easily identify 
the class of variability in each observation. \correc{16~s resolution light curves
were extracted from standard 2 data between 2 and 18 keV. These were directly 
compared to the 3--13~keV and 18--50~keV light curves extracted from the \integral\ data.}
All these light curves were corrected for background, using the latest PCA background models 
available for bright sources.

\subsection{Ryle Telescope observations}
The observations with the Ryle Telescope followed the scheme described by 
\citet{pooley97}. Observations, of Stokes' I+Q, were interleaved 
with those of a nearby phase calibrator (B1920+154), and the flux-density 
scale set by reference to 3C48 and 3C286, and is believed to be consistent 
with that defined by \citet{baars77}. The data are sampled every 8 s,
and 5-min averages displayed in this paper. Fig. \ref{fig:lite} shows the 15 GHz 
long term light curve of \grs.

\section{Accretion ejection links, and the presence of X-ray cycles}
Although the classification of the X-ray classes is purely phenomenological,
we think it is rather important  since it helps when referring to a given observation. 
Furthermore the succession of classes, their relation with the radio behavior,
and the pattern of state transition through a single class probably hides  
rich physical phenomena related to the accretion and ejection mechanisms. Hence, 
we feel it is important to report the identification of all classes and the possible 
transitions we observed during our campaign. Since it is not the core of our paper, 
however, this is given in the Appendix. \correc{The classes of variability 
identified during each observation are reported in Table~\ref{tab:Listsource}.}
In the following we focus on the accretion ejection
links of Obs. 1, 2, 4, and 5  \correc{with particular emphasis on the 
intervals that} respectively belong to classes $\nu$, $\lambda$, $\chi$, and 
$\beta$ during the times we have simultaneous radio and X-ray coverages (see Appendix). 

\subsection{Observation 1: class $\nu$}
As can be seen on Fig. \ref{fig:Oct04} on two occasions, at least, a sequence of 
$>100$~s long X-ray dip ended by an X-ray spike (hereafter cycles) were followed 
by radio flares indicative of ejection of material \citep[e.g.][]{mirabel98}. 
\correc{The cycles are all well defined above 18~keV (Fig.~\ref{fig:Oct04}), with almost 
identical patterns as in the  soft X-rays although the dip and spike have an amplitude
less marked. A notable difference with the soft X-ray light curve, however, 
is the presence of a short dip close to the spike.}
The shapes of the two radio flares are  similar (Fig. \ref{fig:Oct04}).  
To estimate their true amplitude 
above a  noise level, we estimated the variance of the radio light curve when in the non-flare 
intervals. The typical RMS was 3.0 mJy. 
The flares reached maxima (calculated from the light curve with a temporal resolution of 5~min) of $62\pm3$ mJy 
and $60\pm3$ mJy on MJDs 53296.76 and 53296.83 respectively. \correc{Radio flares have been 
seen to occur as a response to each sequences of X-ray dips-spikes in this particular class 
\citep{klein02}.} \\

\subsection{Observation 2: class $\lambda$}
As can be seen on Fig. \ref{fig:Nov04}, and Fig. \ref{fig:zoomLambda}, on one occasion, 
an X-ray cycle was followed by a radio flare indicative of an ejection event 
\citep[e.g.][]{mirabel98}. 
The shape of the flare is rather symmetric. To estimate 
the statistical significance of the flare, we calculated the variance of the first part of 
the radio light curve. From this we can calculate a typical RMS of 2.3 mJy. The flare thus 
had an amplitude of $40\pm2.3$ mJy (in the 5min bins light curve Fig. 
\ref{fig:zoomLambda}). This observation of a radio flare during a class $\lambda$ is the first 
ever reported. \correc{As for the other classes with cycles, e.g. class $\nu$, the dips of class 
$\lambda$ are known to be spectrally hard \citep{belloni00}. This is obvious in the SR 
of Fig.~\ref{fig:zoomLambda}. The dip, here mainly visible at soft X-rays, ended 
with a soft X-ray spike, corresponding to a sudden decrease of the $>18$~keV emission 
(Fig.~\ref{fig:zoomLambda}).} \\
\indent Interestingly, two small radio flares seem to have occured prior to the main 
one around MJDs 53324.68 and 53324.75 (Fig. \ref{fig:zoomLambda}). They had respective 
amplitudes of 15.1 and 11.3 mJy. The flares were then significant at respectively 6.5 and 
4.9 $\sigma$. Deep inspection of the X-ray light curve shows that 
they followed short X-ray dips that occured around MJD 53324.66 and MJD 53324.72 
(Fig. \ref{fig:zoomLambda}). Unfortunately, during the first X-ray dip, for which the 
decrease of the 3--13~keV count rate is clearly visible (Fig. \ref{fig:zoomLambda}), 
\integral\ did a slew between two subsequent pointings that prevented a full coverage 
of this interval. We can just set a lower limit on the duration of the dip of 100s. 
In the second case, the delay between the return to a high degree of variability at 
X-ray energies and the peak of the radio flare was 0.15 hours. \correc{Although the 
cycle was shorter than during the long one preceding the 40~mJy flare, a similar
sequence of event can be seen here. The dip is really apparent below 13~keV and 
is spectrally hard as illustrated by the SR on Fig.~\ref{fig:zoomLambda}. The cycle 
ended with a return to a high level at soft X-rays associated with a slight decrease 
of the hard X-ray emission. As in the previous cases the cessation of the dip 
manifested by a peak (although of much smaller amplitude than in the main cycle) 
in the SR Fig.~\ref{fig:zoomLambda}.}

\subsection{Observation 4: class $\chi$}
Contrary to the previous classes, \grs\ showed a relatively steady emission at 
all wavelengths, although some variations, in particular near the end of the observation
are visible (Fig. \ref{fig:April04}). In the radio domain the source had a 
mean flux of 44.9 mJy with a typical RMS of 3.0 mJy (calculated from the radio light curve 
with a binning of 5 min) from MJD 53473.10 to MJD 54473.24. It then slowly increased during 
$\sim4800$ s to reach a mean flux of 70.4 mJy, with a RMS of 4.4 mJy from MJD 53473.30 to 
53473.42. \correc{The mean 3--13~keV JEM-X was 115.7$\pm$6.8~cts/s from 
MJD 53473.10 to 54473.24  and 107.3$\pm$6.5~cts/s later (MJD 53473.30--53473.42). 
The mean 18--50~keV ISGRI count rates was 55.3$\pm$2.8~cts/s from 
MJD 53473.10 to 54473.24 and 58.4$\pm$2.7~cts/s after (MJD 53473.30--53473.42).}\\
\indent Near the end of the observation (MJD 53474.4), the radio flux was still around the same 
value as before the radio coverage was stopped (Fig. \ref{fig:April04}). It decreased for a short 
time to $\sim 59$~mJy and increased to $\sim 110$~mJy after MJD 53472.2. Note that an 
inspection of the radio light curve \correc{the following days}, showed that the radio flux 
was at approximately the same level (120-130 mJy) on MJD 53477, and had decreased 
to $\sim 60$ mJy on MJD 53478. This may suggest that the radio
emitter, the jet, has persisted over a long period, although it showed slight variations in 
its strength.

\subsection{Observation 5: class $\mu$ and $\beta$}
During this observation, again on two occasions, the X-ray dips were followed by radio flares with 
symmetric shapes (Fig. \ref{fig:May05}). The radio flares reached their maxima on MJDs 53504.248 
and 53504.298 respectively, with absolute peak fluxes of $58.5$ and $68.3$ mJy (when using 5 min 
bins to estimate them).
Unlike the two other classes, they, however, sat on top of a non-zero radio flux. The mean radio 
flux prior to the two flares was 35.2 mJy with an RMS of 3.4 mJy. The net amplitude of the flares above
the continuum emission was therefore $23.3\pm3.4$ mJy and $33.1\pm3.4$ mJy, for the first and second 
flares respectively. The time delays between the X-ray spikes half way through the dips (the triggers 
of the ejections, \citet{mirabel98,chaty98}), and the radio maxima were  $\sim 0.31$ hour and 
$\sim 0.34$ hour respectively. The presence of a relatively high radio flux prior to the 
ejections, when \grs\ was showing a high level of very variable X-ray emission 
(Fig. \ref{fig:May05}), is quite interesting. This kind of steady radio flux is usually indicative 
of the presence of a compact jet. The latter is, however, usually seen when the source shows 
a steady X-ray flux dominated by hard X-ray emission, its spectrum being, then, indicative of a 
hard-intermediate state (state C of \citet{belloni00}), similar to the one seen during Obs. 4.
\citet{klein02} also report the presence of radio emission during class $\mu$ observation with 
no oscillations of the radio flux. This could indicate that the cycles in class $\mu$ are too 
fast to be detected at radio wavelength. We should also note that the radio non-zero continuum seems 
to be the tail or exponential decay of a major flare that occured some days before the observation
as can be seen in Fig.~\ref{fig:lite}.
\correc{As in classes $\nu$ and $\lambda$, the cycles are also well defined above 18~keV. In 
particular, a dip can clearly be seen at hard X-ray energies during each cycle.}

\section{Discussion}
We have presented the results of 2 years of simultaneous monitoring campaigns on the Galactic 
microquasar \grs\ made with several instruments. This has  allowed  
us to follow the behavior of the source 
from radio wavelengths to hard X-ray energies. The \integral\ observatory, thanks to its 
large field of view, has allowed us to quickly monitor the behavior of sources that are 
in the vicinity of \grs. \\
\indent In the case of \grs, we first classified the X-ray classes of 
variability of the source (Appendix), and saw that \grs\ could undergo 
transitions through many classes of variability on short time scales 
(few hundreds seconds, Appendix). \correc{We report here the following 
specific and direct transitions: $\nu\leftrightarrow\rho$ (Obs. 1), 
$\mu\leftrightarrow\lambda$ and $\mu\rightarrow\delta$ (Obs. 2), 
$\mu\rightarrow\beta$ (Obs. 5), $\phi\rightarrow\theta\rightarrow\delta$ (Obs. 7), 
$\chi\leftrightarrow\theta$ (Obs. 8 \& 9, Ueda et al. 2006), possibly 
$\delta(?)\rightarrow\mu\rightarrow\beta$ (Obs. 10), and 
$\delta\rightarrow\mu\rightarrow\beta$ (Obs. 11)}.
We then focused more particularly on 3 observations showing the 
occurrences of X-ray cycles followed by radio flares. In a fourth observation
the source was in its steady ``hard'' state accompanied by the presence of a persistent 
radio emission.\\

\subsection{Steady radio and X-ray emission}
\indent During Obs. 4, the persistent radio emission can be safely attributed to a steady 
compact jet, although the lack of simultaneous coverage at other radio wavelengths prevents us 
to precisely estimate the radio spectral index. \grs\ is in a class $\chi$ also known as 
being its ``hard'' state (in fact it rather corresponds to a hard-intermediate state in 
the recent classification of \citet{homan06}), a state during which strong radio emission 
associated to a compact jet has been seen in the past \citep[e.g.][]{fuchs03}. During this 
particular observation, the radio flux increased \correc{by a factor of 1.57} in a rather short 
time (less than an hour), \correc{while the 3--13~keV and  
the 18--50~keV count rates remained constant at the 1-$\sigma$ level.} 
\correc{The fact that the X-ray count rates do not 
follow the evolution of the radio flux may indicate that the jet has no influence 
on the X-ray emission at all. However, \grs\ is a source known to follow the radio to X-ray 
correlation seen in many microquasars \citep{corbel03,gallo06}, which has been 
widely interpreted as evidence for an influence of the jet at X-ray energies.  A way to reconcile 
our observation to the latter interpretation is to suppose that the  X-ray count rates 
come from ``competing'' media emitting in the same energy ranges, 
e.g. a disk, and/or a standard corona and/or the jet. A model-dependent 
analysis is, however, beyond the scope of this paper.}\\

\subsection{On the generalization of the radio to X-ray connection}
\begin{figure}
\figurenum{7}
\epsscale{1}
\caption{Zooms on portions of Obs. 2, 3 and 5 (from left to right). {\bf{Top panel:}} 
JEM-X 3--13 keV light curves with for each observations the 3 intervals with specific 
properties discussed in the text. {\bf{Middle panel:}} 18--50~keV ISGRI light curves. 
{\bf{Bottom panel}}: 3--13~keV/18--50~keV SR. Note that different vertical 
scales of each individual plots.}
\plotone{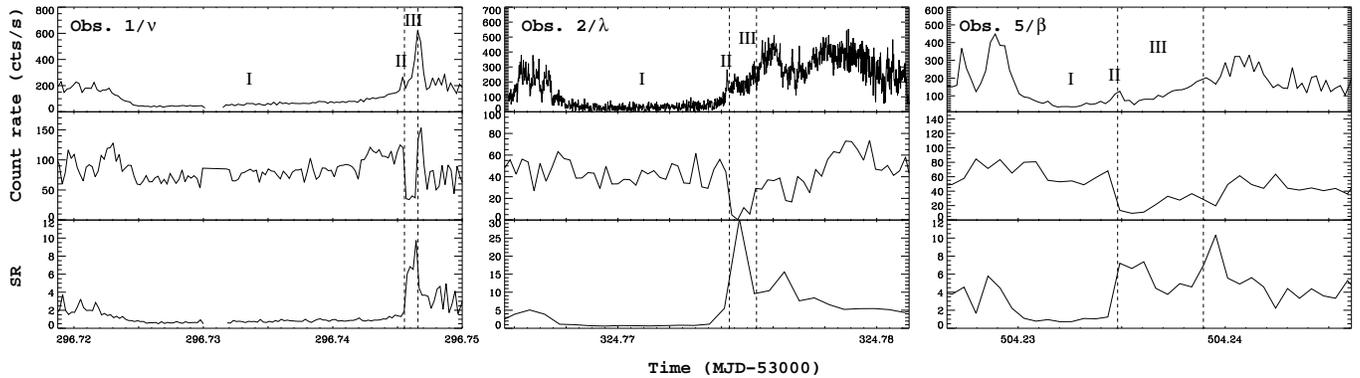}
\label{fig:zooms}
\end{figure}

\indent Focusing on the observations showing X-ray cycles, the observations of links between 
the X-ray cycles in class $\nu$ and $\beta$ confirm 
previous findings that such events seem to be generic in those classes 
\citep{pooley97,mirabel98,klein02}. 
The observation of radio ejection in the $\lambda$ class of variability is, 
however, the first ever reported. This may suggest a generalization of  
the fact that small amplitude ejections in GRS 1915+105 
always occur as a response to X-ray cycles providing the X-ray hard dip is long enough 
\citep{klein02}. The observation by \citet{feroci99} of an X-ray dip not followed by a 
radio flare during a {\it{BeppoSax}} observation may be in contradiction with 
this generalization.  \\
\indent \correc{In fact, although some obvious morphological differences exist between the 
different classes, in term of SR and colors they all undergo similar evolutions 
\citep[][and Fig.\ref{fig:zooms}]{belloni00}.}  The cycles always begin with a transition 
to a low flux below 13 keV, the  X-ray dip (interval I in Fig.\ref{fig:zooms}), associated 
to a relatively bright flux above 18 keV. \correc{The 3--13~keV/18--50~keV SR has then a value
of about 1. This indicates the dip is spectrally hard.}  A short spike (interval II) 
occurs in the 3--13~keV range (\correc{it is very short in class $\lambda$}). This 
spike seems to be the onset of a sudden and fast \correc{change, as the rising part is still 
hard (SR$\sim1$), although it smoothly evolves. After the spike, however, 
a fast and important decrease} of the hard X-ray emission (interval 
III) during which \correc{the SR increases greatly, indicates a much softer state is reached.} 
In \correc{class $\nu$, for example, the 18--50 keV count rate} decreases by a factor $\sim$3.
The evolution of the soft X-ray emission is less dramatic, although it decreases as 
well from interval II to III in all classes, its evolution seems to be the 
continuation of the slow increase seen in each case at the end of interval I (Fig. \ref{fig:zooms}).
\citet{mirabel98, chaty98} identified the spike (II) in class $\beta$ as the trigger 
to the ejection later seen in radio. In all three cases the delay between Int. II seen  
at X-ray energies and the peak of the radio flare is very similar. In class $\nu$ it is respectively 
$\sim 0.31$ hour and $\sim0.29$ for the first and second flares respectively. In class $\lambda$,
it is $\sim0.31$ hour, while in class $\beta$ it is respectively $\sim 0.31$ hour and 
$\sim 0.34$ hour for the first and second flare. 
This similarity may suggest that the same physical mechanism in the three classes give rise to 
the same phenomenon. Hence if the spike (Int.~II) in class $\beta$ is indeed the trigger of 
the ejection, it has the same role in classes $\nu$ and $\lambda$.  
\correc{This model independent approach may suggest that the ejected material is the material 
responsible for the hard X-ray emission prior to the ejection occuring at Int.~II. A similar
interpretation was given by \citet{chaty98} and \citet{rod02b} in the case of class $\beta$ and
$\alpha$ observations.}\\
\indent  The lack of ejection after the X-ray dip reported by \citet{feroci99} can then easily 
be understood. 
In their case the dip in question is not followed by a spike, while an ejection they observe
follows an X-ray spike, possibly following an X-ray dip 
(but missed due to occultation of \grs\ by the Earth). Interestingly, the approximate delay between the 
X-ray spike and the maximum of the radio flare in this observation (taken in 1996, \citet{feroci99}) 
is $\sim0.28$ hour, hence very similar to what we obtain in all our cases. \correc{In addition, 
the X-ray dip discussed in \citet{feroci99} seems much softer than the remaining intervals 
of their observation \citep[see Table 1 of][]{feroci99}, as its photon index 
is the softest of the sequence}. Putting everything together, 
it seems, then, that we can generalize the following: plasmoid ejections {\it{always}} occur as 
response to an X-ray sequence composed of {\it{a longer than 
100s \correc{spectrally hard} X-ray dip terminated by a short X-ray spike}}, the latter being 
the trigger of the ejection. \correc{This possible generalization is even re-inforced by the 
observation of smaller amplitude ejections following short cycles during class $\lambda$. Although
the shorter duration of the events prevented us to obtain the same details as in the main cycles, it 
is obvious from Fig.~\ref{fig:zoomLambda} that \grs\ undergoes similar evolution during one of the 
shorter dips. In particular, the dip is spectrally hard (SR$\sim1$), while it ends with a 
small 3--13~keV spike, also visible in the SR which increases to about 9, and therefore
marks a transition to a much softer state.}\\

\subsection{A link between the radio amplitude and the duration of the X-ray dip?}
\begin{figure}
\epsscale{0.5}
\figurenum{8}
\caption{Evolution of the amplitude of the radio flares vs. the duration 
of the preceding X-ray dip. The points 
are from this study and the triangles come from \citet{pooley97}, and \citet{klein02}.
 The  line represents the linear function that best fits the data.}
\plotone{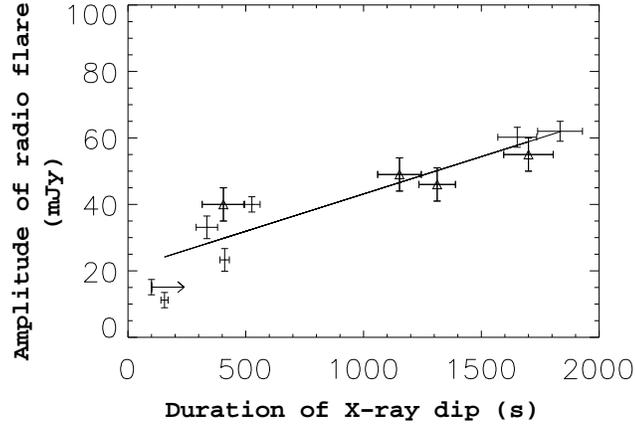}
\label{fig:correl}
\end{figure}

\indent Looking at class $\lambda$ in more detail, we detected the presence of two 
small radio flares each apparently following an X-ray dip of short duration. In order to study 
a possible dependence of the amplitude of the radio flare on the duration of the X-ray dip, 
we estimated the duration of the X-ray dips in each class. Since the dips have 
different shape, and 
 the transition into the dip is quite smooth, the true starting time of the dip is 
quite difficult to estimate. To do so, we take as the starting point of the 
dip the mean time between the highest point immediately preceding the transition to the dip 
and the time at which the bottom of the dip is truly reached. The error is here the time difference 
between the highest  point immediately preceding the transition to the dip 
and the bottom of the dip. 
The end time is taken as the ``foot'' of the X-ray spike which 
renders its identification easier, given that 
the transition into the spike is quite sudden and rapid in all cases. The results are plotted in 
Figure \ref{fig:correl}. A positive correlation between the two quantities is quite obvious.
 The linear Spearman correlation coefficient is 0.93. The best linear fit leads
to $F_{15~{\rm{GHz}}}{\rm{(mJy)}}=0.025\times t{\rm{(s)}} +17.2$. Note, however, 
that a pure linear dependence of the radio amplitude 
vs. the duration of the X-ray dip remains unlikely as \citet{klein02} remarked that for an 
ejection to take place a dip of minimum duration $\sim100$~s is necessary. \\
\indent In order to populate the region around 1000~s, we searched the literature 
for simultaneous radio (15 GHz) and X-ray data showing the occurrences of cycles, and ejections. 
Some clear examples are given in \citet{pooley97} and
\citet{klein02}, for a total of 8 cycles occuring around MJDs 50381.6, 50698.8, and 51342.1. 
However, in 4 cycles (on MJDs 50698.76 50698.78, 51342.06 51342.12)
the radio flares sat on top of non-zero (30 to 60 mJy) radio continuum 
which renders the estimate of the radio amplitude very uncertain. Note that the 
binning of 32 s presented in \citet{klein02} is also another source 
of uncertainty. These four cycles are not included here. The radio flares around 
MJD~50381.6 \citep{pooley97} 
also sat on a non zero continuum, but the latter is quite weak ($\sim 10$ mJy), and 
therefore the amplitude of the radio flare can be estimated more accurately. Finally 
the last radio flare, occurs on MJD 50698.83 just after a period of $\sim$0~mJy level. 
These four additional cycles are added in Fig. \ref{fig:correl},
and represented as triangles. With these new points, the general tendency remains the same. 
The linear Spearman correlation coefficient is 0.90, and the best linear fit leads 
to $F_{15~\rm{GHz}}{\rm{(mJy)}}=0.022\times t{\rm{(s)}} +20.7$. \\
\indent \correc{We also searched for other possible correlations between some properties of 
the cycles, and the amplitude of the radio flares, as, e.g., the amplitude of the spike (with
respect to the bottom of the preceding dip), the amplitude of the variations of the SR after 
the spike, or the time delay between the X-ray spike and the peak of the radio flare. 
We do not find any obvious correlations between any of these quantities. 
The  maximum variation for the SR is seen 
during the main $\lambda$ cycle, while it corresponds to a rather low amplitude ejection. No
correlation is either found between the amplitude of the spike and the amplitude of the radio flare.
The maximum amplitude of the spike occurs during the main cycle of class $\lambda$ observation, 
for a relatively low amplitude radio flare. The spike with the minimum amplitude occurs during 
class $\beta$ observation for a radio flare of  similar amplitude to that of class $\lambda$. 
}\\
\indent The  correlation of the radio amplitude 
with the duration of the X-ray dip brings interesting possibilities regarding the 
accretion-ejection links. This may indicate, for example, that during the X-ray dip 
energy and matter, later used to power the ejection, are accumulated. The longer 
the X-ray dip, the more energy and/or matter are accumulated, and then the  
higher is the radio amplitude. \correc{This is compatible with the fact that  
the ejected material is the matter responsible for the hard X-ray emission 
suggested by the sudden decrease of the 18--50~keV emission at the spike. In that case,
a longer duration of the dip would indicate that more matter, later ejected, is 
accumulated during the dip.} \\
\indent Another possible explanation (not exclusive with the previous one) relies on 
the so-called `Magnetic Floods' scenario/model \citep{tagger04}. 
This model was recently proposed to account for the observed accretion-ejection 
and quasi-periodic variability properties, during the cycles observed in a 
class-$\beta$ observation. In this scenario, during the X-ray dip 
an `Accretion-Ejection Instability' (AEI, \citet{tagger99}) develops 
(and replaces the Magneto-Rotational Instability (MRI) thought to occur 
during the preceding luminous soft X-ray state), because poloidal magnetic 
field advected with the matter has accumulated in the inner region of the disk. 
If the magnetic configuration is favorable, a sudden reconnection event 
(producing the spike) can occur between magnetic fields of opposite polarities 
in the disk and in the magnetospheric structure of the black hole. This would then lead to the 
dissipation of the accumulated field in the inner regions, leading to the ejection 
(of the corona), and return to the MRI \citep{tagger04}. This scenario had some success in 
explaining all observational signatures (including QPOs) seen during the particular class 
$\beta$ these authors dealt with. This interpretation is also compatible with the observed
behavior we present here, and, in particular, the possible correlation between the amplitude of 
the radio flare and the duration of the dip (Fig. \ref{fig:correl}). In that case, the longer the dip, 
the more magnetic flux can be accumulated in the inner region, and again the higher is 
the available energetic reservoir used in energizing the ejection. This interpretation is however 
only tentative, and should not hide the fact that other models may explain these observations. However, 
the generalization of the X-ray cycles to radio ejections and the relation between the duration 
of the dip and the amplitude of the ejection bring strong constraints on any attempt to model 
the accretion-ejection behavior in \grs\ and other microquasars. Note that the possible correlation 
we found here clearly needs to be confirmed through systematic inspection of simultaneous radio 
and X-ray data, but also adding different frequencies (e.g. IR data).\\

\begin{acknowledgements}
J.R. would like to thank S. Chaty for useful discussions, C.A. Oxborrow for invaluable help 
with the ISGRI and JEM-X data reduction and calibration, and A. Gros, A. Sauvageon and 
N. Produit on specific aspect of the \integral\ software.
J.R. also acknowledges E. Kuulkers, and more generally
the \integral\ and \rxte\ planners for their great help in scheduling the observations. D.C.H. 
gratefully acknowledges a Fellowship from the Finnish Academy. AP acknowledges the Italian 
Space Agency financial and programmatic support via contract ASI/INAF I/023/05/0. The authors thanks 
the referee for his carefull reading and valuable report which helped to greatly improve the quality of 
this paper.\\
Based on observations with \integral, an ESA mission with instruments and science data centre funded by ESA 
member states (especially the PI countries: Denmark, France, Germany, Italy, Switzerland, Spain), 
Czech Republic and Poland, and with the participation of Russia and the USA.
This research has made use of data obtained through the High Energy 
Astrophysics Science Archive Center Online Service, provided by the NASA/
Goddard Space Flight Center.
\end{acknowledgements}

\bibliography{ms}

\appendix
\section{Identification of the X-ray classes}
\begin{figure}
\epsscale{1}
\figurenum{2}
\caption{{\bf{Left:}} Light curves of \grs\ during Obs. 1: a$)$ RT at 
15 GHz, b) JEM-X 3--13 keV \correc{binned at 20~s}, c) 
ISGRI \correc{18--50 keV binned at 50~s}, d)  \rxte/PCA \correc{2--18 keV binned at 16~s}. 
{\bf{Right:}} CC diagrams (upper panels)
and \rxte\ light curves (lower panels) of 2 sub-intervals from Obs. 1, showing 
occurrences of class $\rho$ (left) and class $\nu$ (right).}
\plottwo{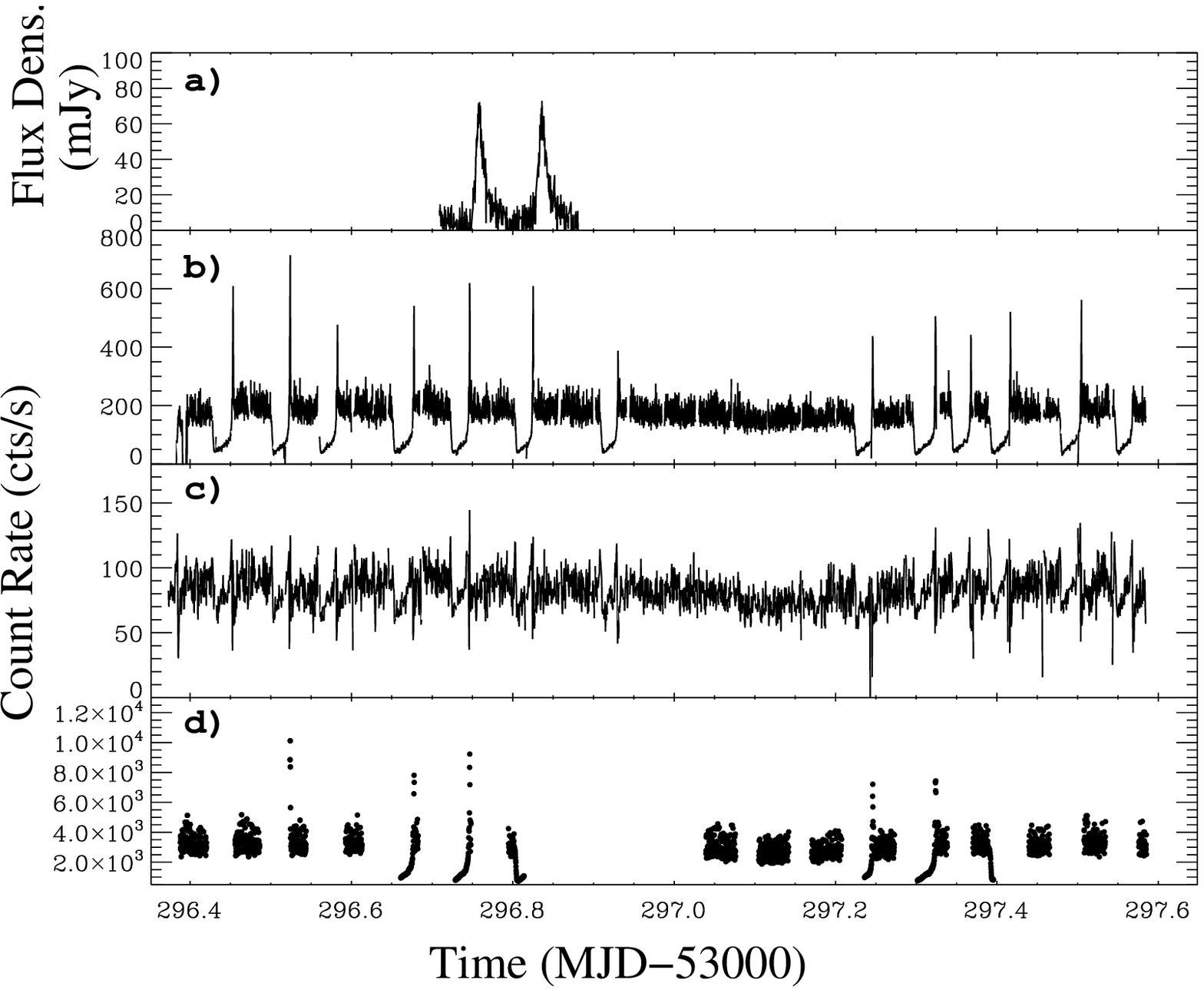}{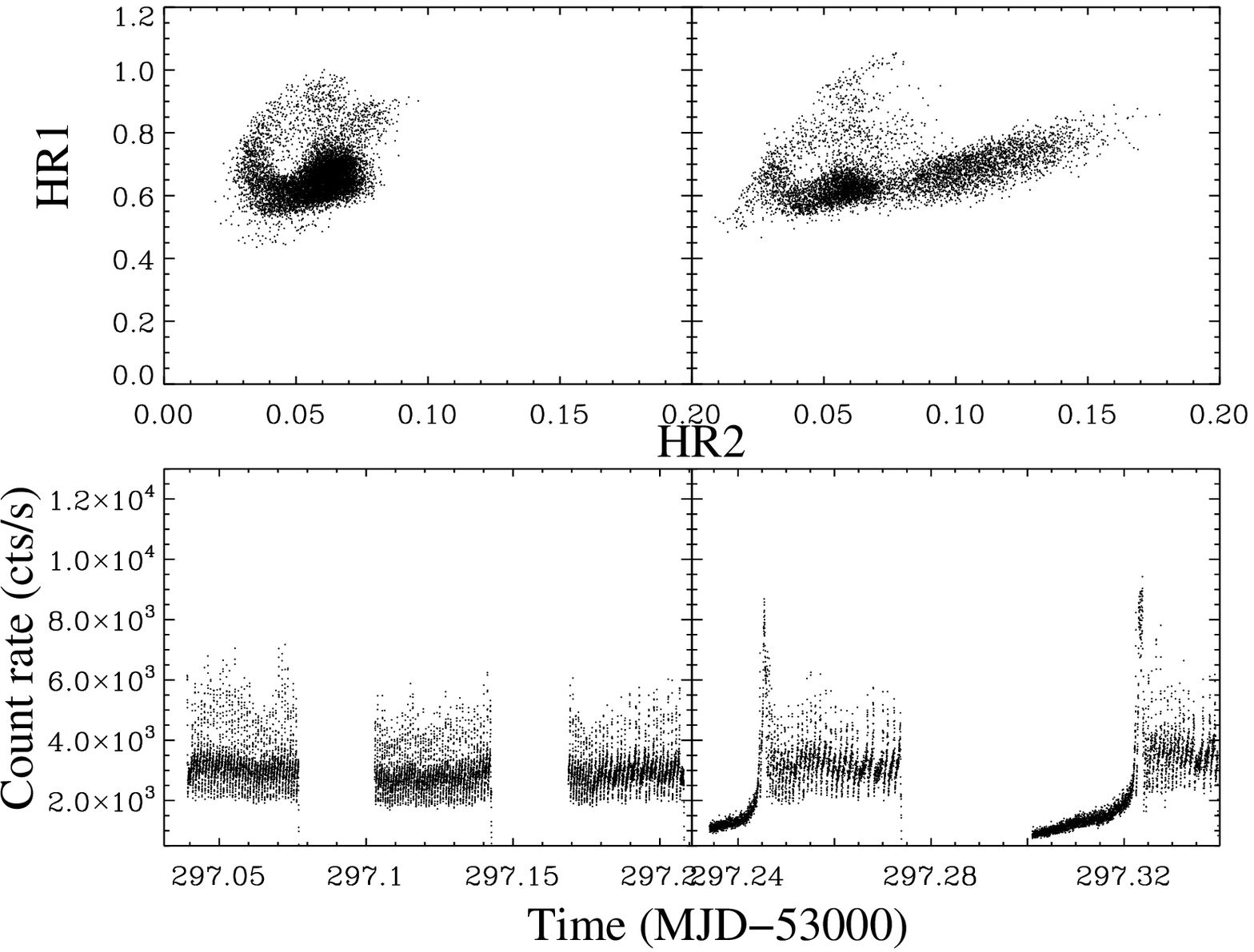}
\label{fig:Oct04}
\end{figure}

\begin{figure}
\figurenum{3}
\epsscale{1}
\caption{{\bf{Left:}} \correc{Same as Fig.~\ref{fig:Oct04} for Obs. 2.} {\bf{Right:}}  CC diagrams
 (upper panels) and \rxte\ light curves (lower panels) of 3 sub-intervals from Obs. 2, 
showing occurrences of class $\mu$/$\kappa$ (left), class $\lambda$ (middle) and class $\gamma$ 
(right).}
\plottwo{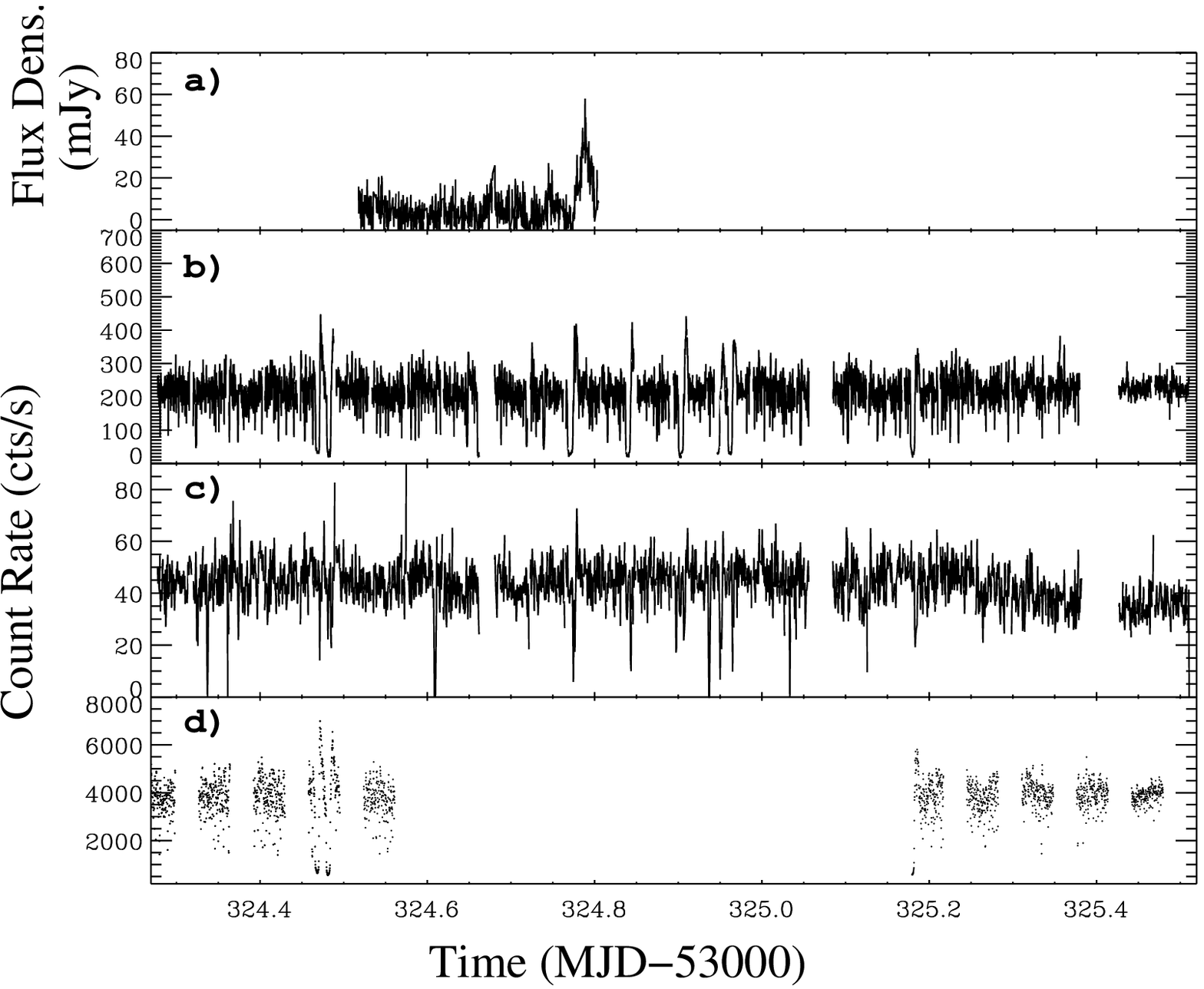}{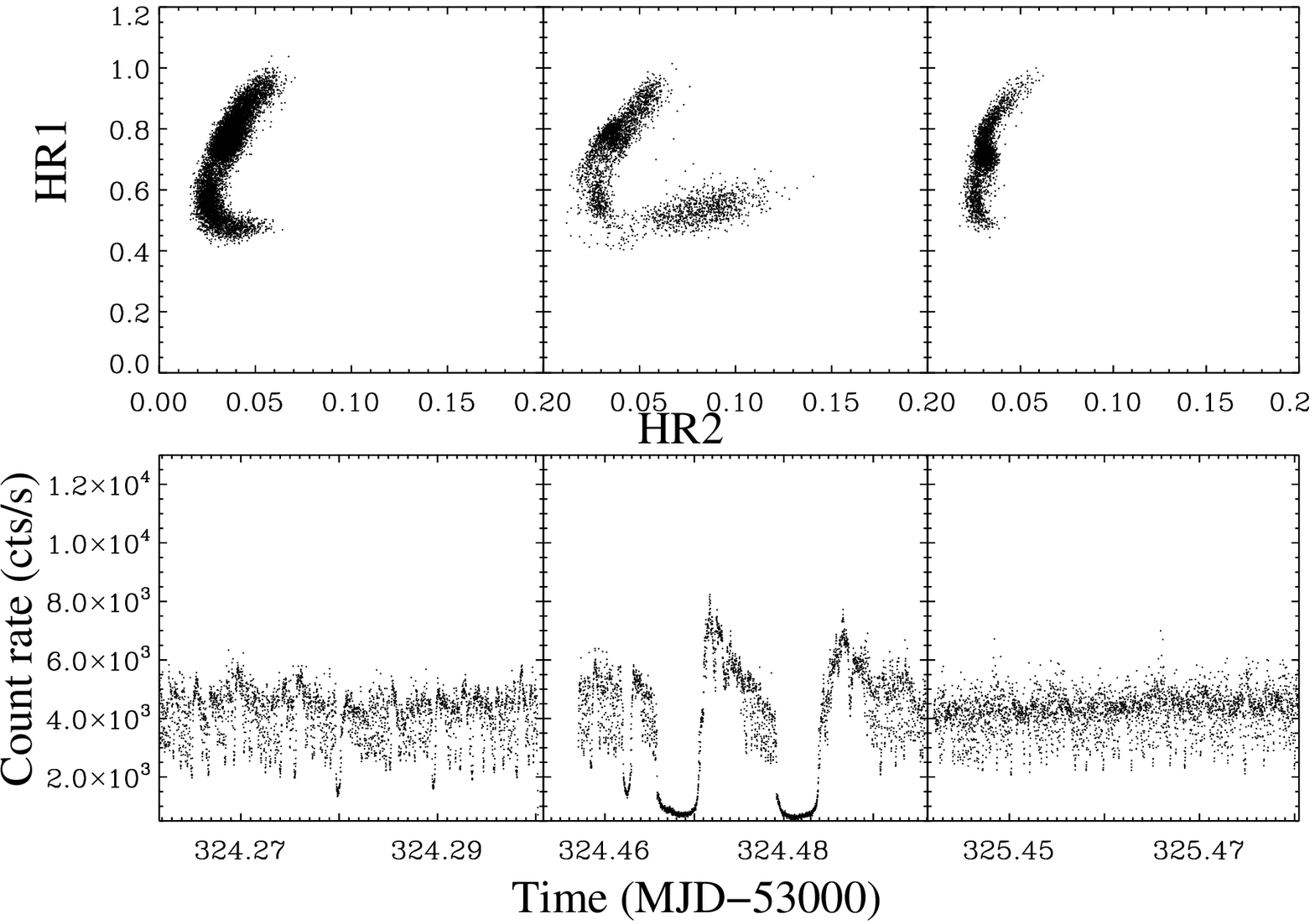}
\label{fig:Nov04}
\end{figure}

\begin{figure}
\figurenum{4}
\epsscale{0.5}
\caption{Zoom on a portion of Obs. 2 showing in the top panel three radio flares 
(indicated by the arrows). \correc{The other panels show, from top to bottom, the JEM-X 3--13~keV 
count rate, the ISGRI 18--50~keV count rate, and the 3--13~keV/18--50~keV SR. The box delimited with 
dashes shows the short cycle preceding the second radio flare.}}
\plotone{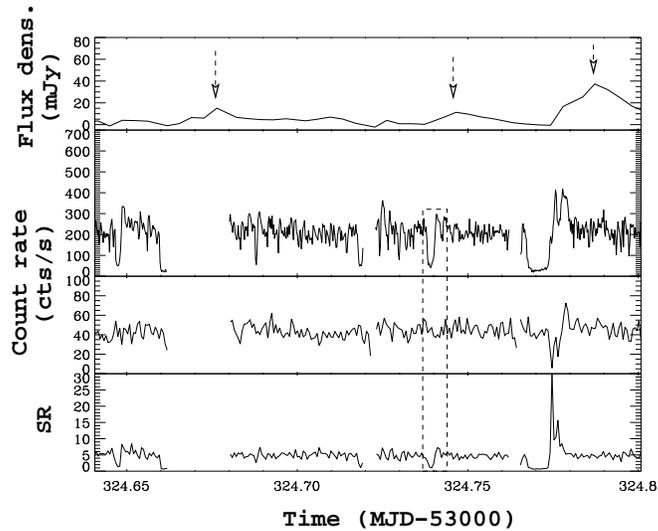}
\label{fig:zoomLambda}
\end{figure}

\begin{figure}
\figurenum{5}
\epsscale{1}
\caption{{\bf{Left:}} \correc{Light curves of \grs\ during Obs. 4: a$)$ RT at 
15 GHz, b) JEM-X 3--13 keV binned at 50~s, c) 
ISGRI 18--50 keV binned at 200~s, d) \rxte/PCA 2--18~keV binned at 16~s}. 
{\bf{Right:}}  CC diagram
 (upper panel) and \rxte\ light curve (lower panel) of 1 sub-interval from Obs. 4 
showing the source was in class $\chi$.}
\plottwo{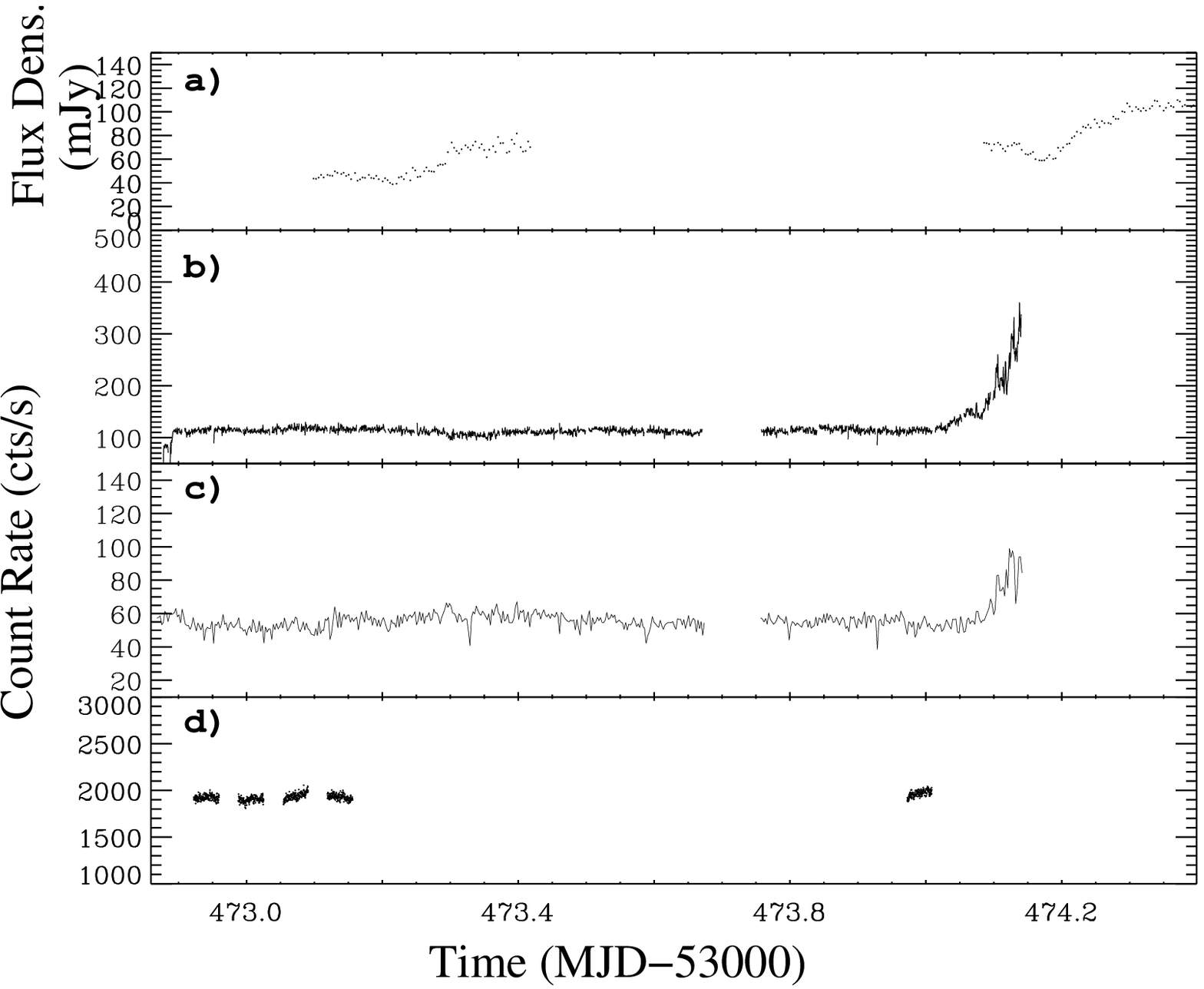}{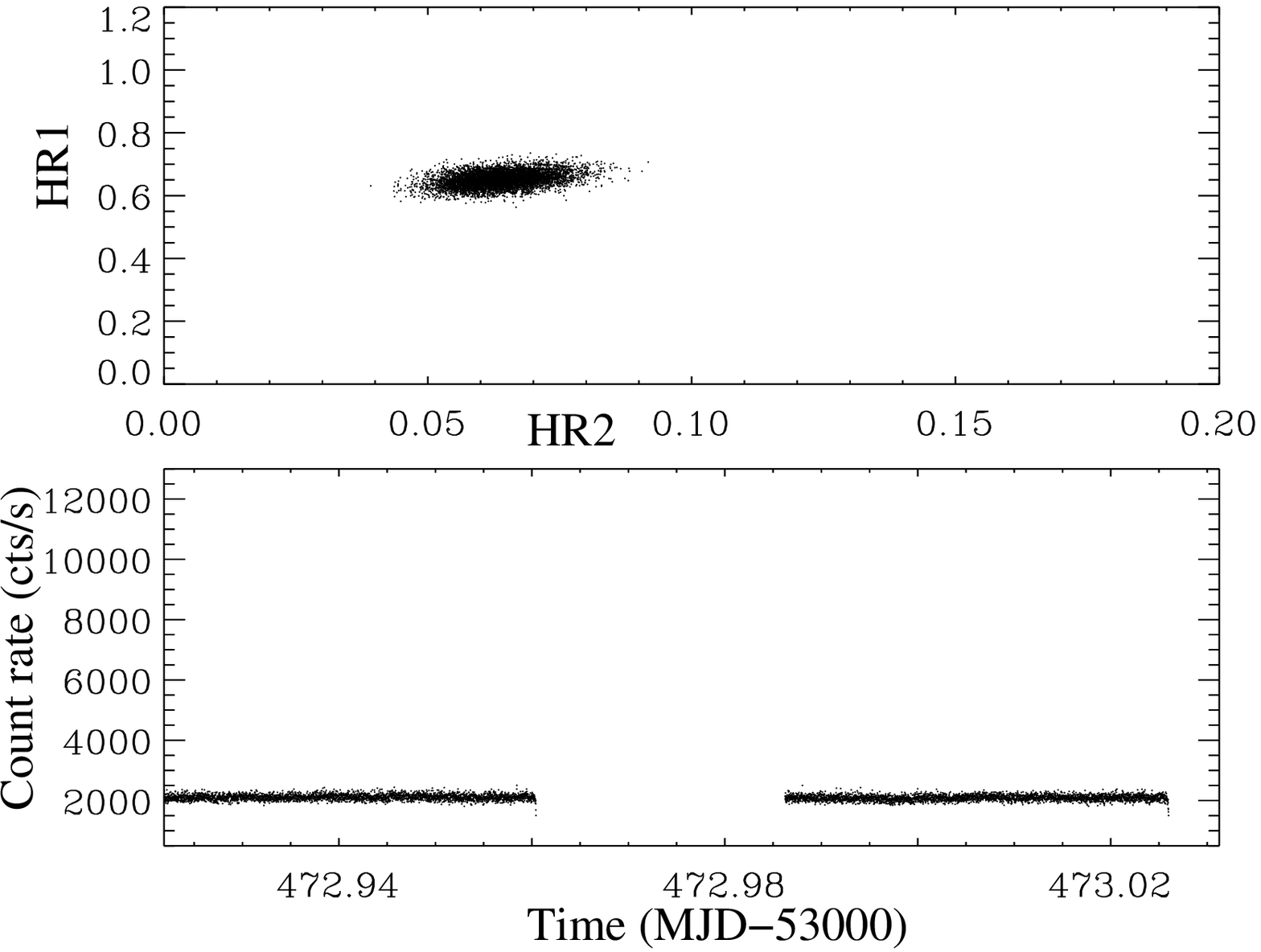}
\label{fig:April04}
\end{figure}

\begin{figure}
\figurenum{6}
\epsscale{1}
\caption{{\bf{Left:}} \correc{Same as Fig.~\ref{fig:Oct04} for Obs. 5.} {\bf{Right:}}  CC diagrams
 (upper panels) and \rxte\ light curves (lower panels) of 2 sub-interval from Obs. 5 showing 
occurrences of class $\mu$ and $\beta$.}
\plottwo{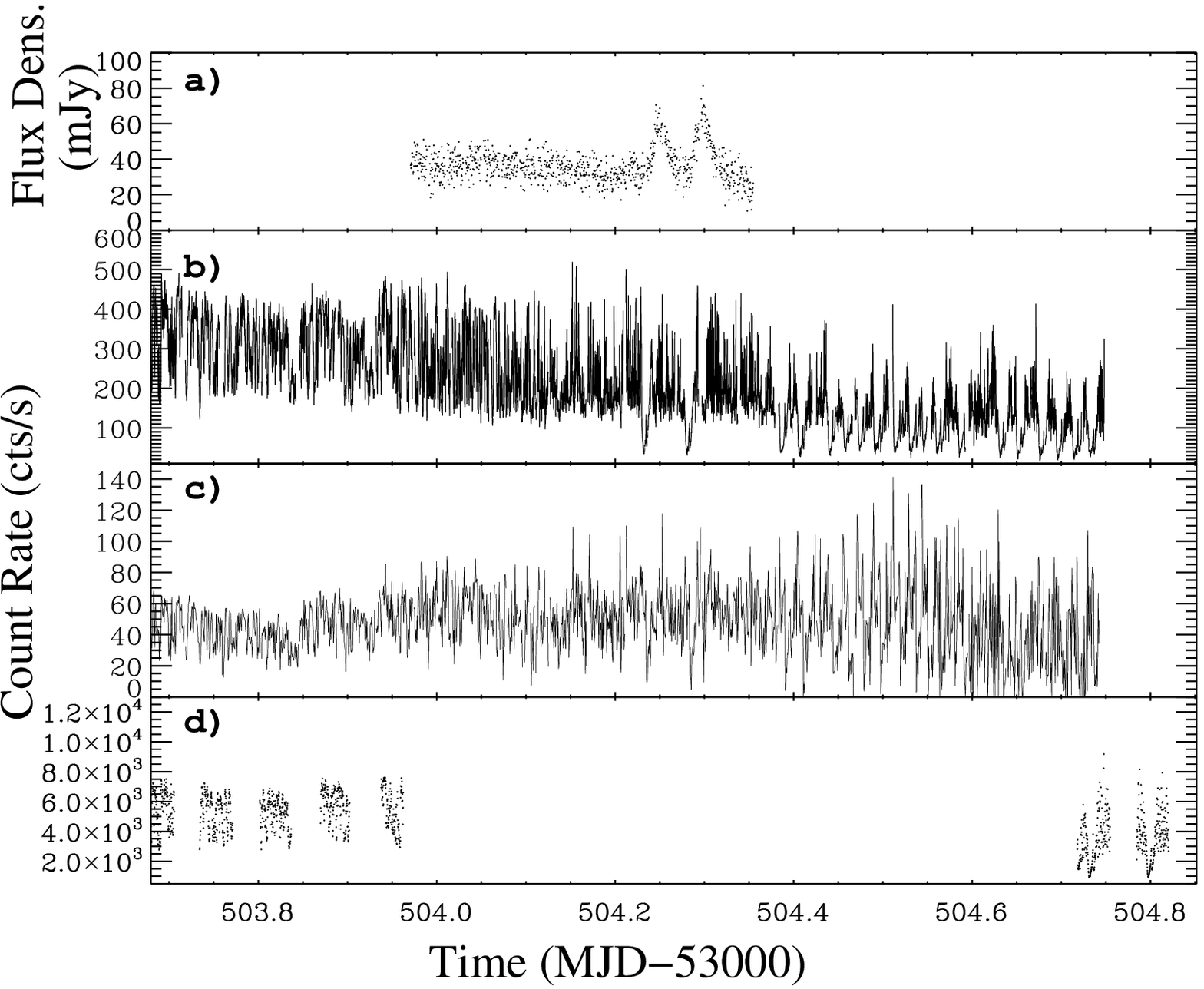}{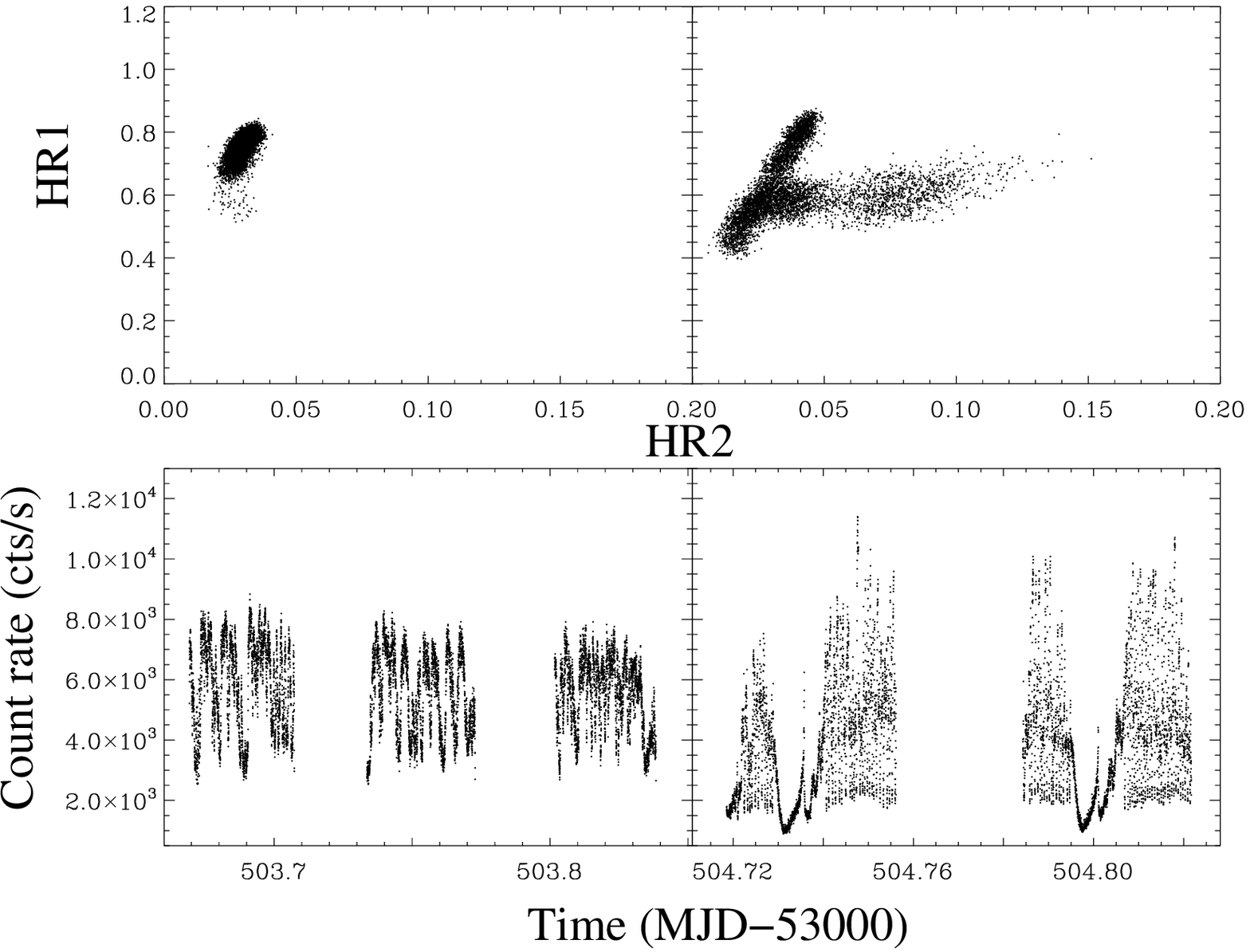}
\label{fig:May05}
\end{figure}

\begin{figure}
\figurenum{10}
\caption{{\bf{Left:}} Light curves of \grs\ during Obs. 7: a$)$ RT at 15 GHz, 
b) JEM-X 3--13 keV binned at 20~s, c) ISGRI 18--50 keV, binned at 100~s. {\bf{Right:}} CC diagrams
 (upper panels) and \rxte\ light curves (lower panels) of 2 sub-interval from Obs. 7 
\correc{showing the source was in class $\phi$ at the beginning and class $\delta$ in the end. 
Note that between those two classes \grs\ transited through a class $\theta$.}}
\plottwo{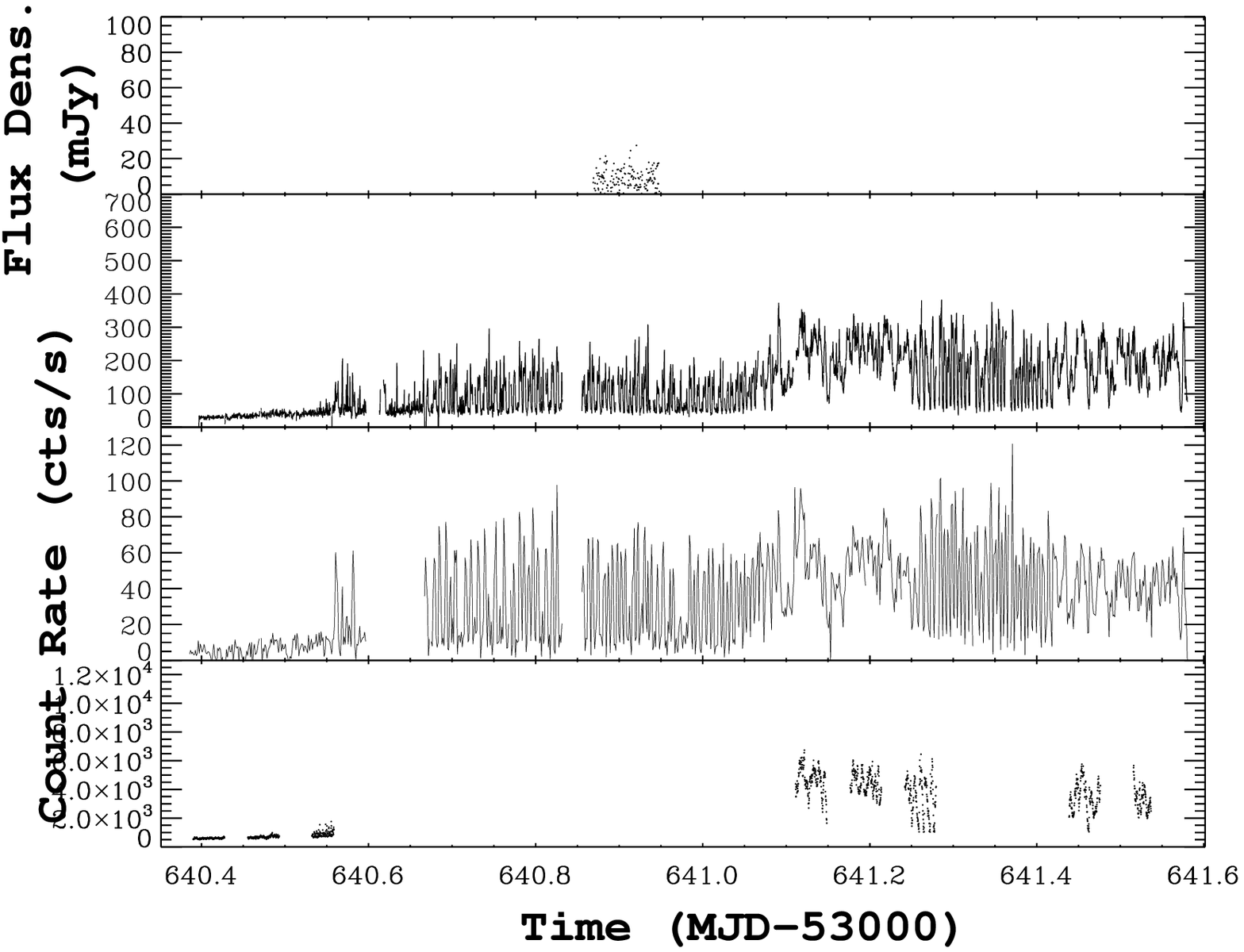}{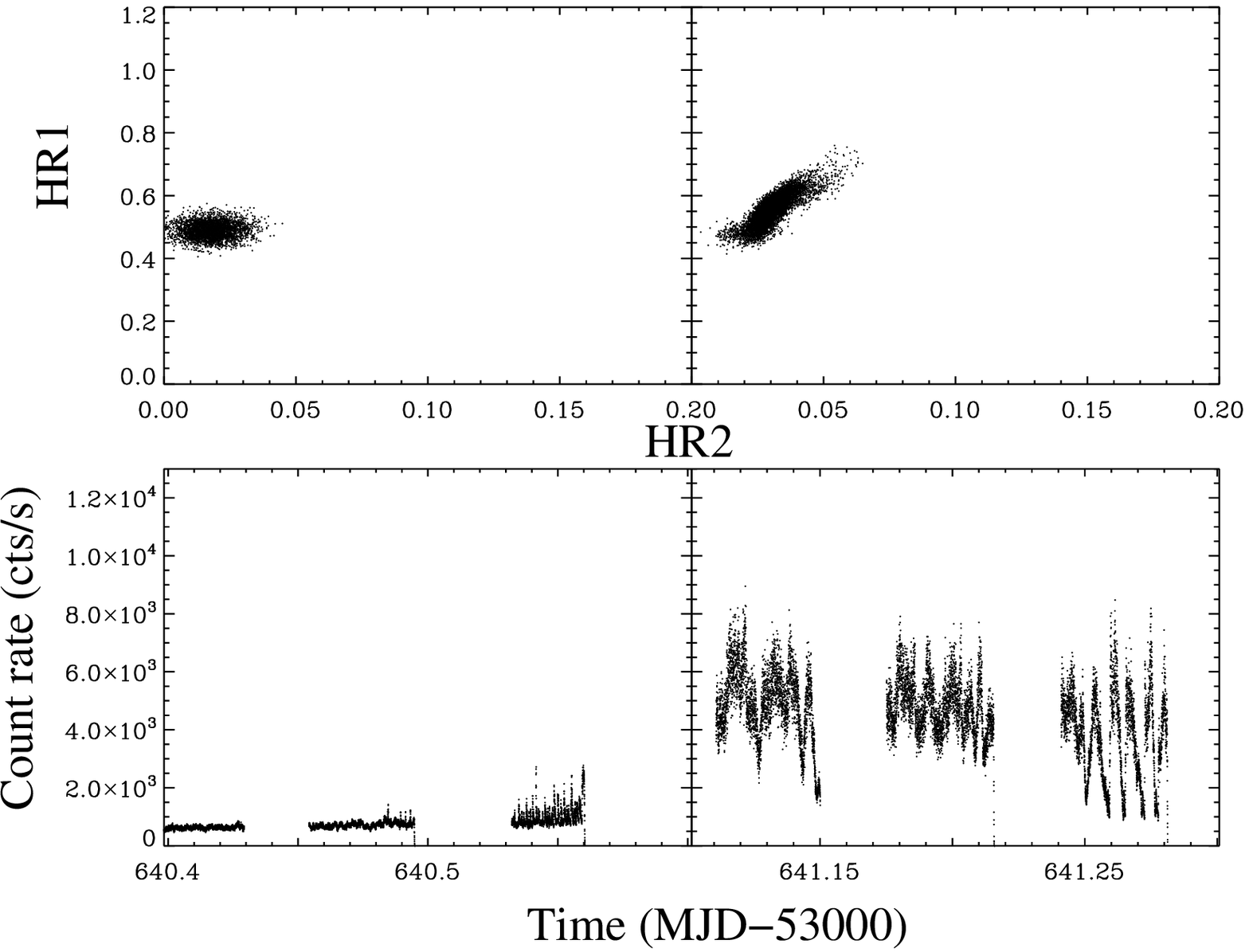}
\label{fig:Sept05}
\end{figure}

\begin{figure}
\epsscale{0.5}
\figurenum{11}
\caption{\correc{Zoom on a  sub-interval from Obs. 7. From top to bottom the 
plots respectively represent the 
JEM-X 3--13 keV , ISGRI 18--50 keV light curves and the 3--13~keV/18--50~keV SR. Both light curves
show ``M-shape'' patterns typical of class $\theta$.}}
\plotone{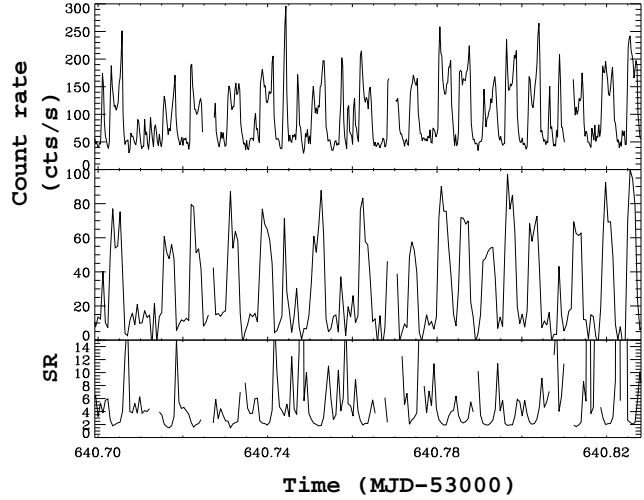}
\label{fig:Sep05Theta}
\end{figure}

\begin{figure}
\epsscale{1}
\figurenum{12}
\caption{\correc{{\bf{Left: }}Light curves of \grs\ during Obs. 8 and 9: a$)$ RT at 15 GHz, 
b) JEM-X 3--13 keV binned at 20~s, c) ISGRI 18--50 keV  
binned at 100~s. The source starts in class $\chi$ in the first part, transits 
to class $\theta$ (not visible here, see Ueda et al. 2006) and transits back in 
a likely class $\chi$ in the second interval visible 
here. {\bf{Right: }} Zoom on two sub-intervals from Obs. 8 (left panel) and 
Obs. 9 (right panel). From top to bottom the plots respectively represent the 
JEM-X 3--13 keV , ISGRI 18--50 keV light curves and the 3--13~keV/18--50~keV SR.}}
\plottwo{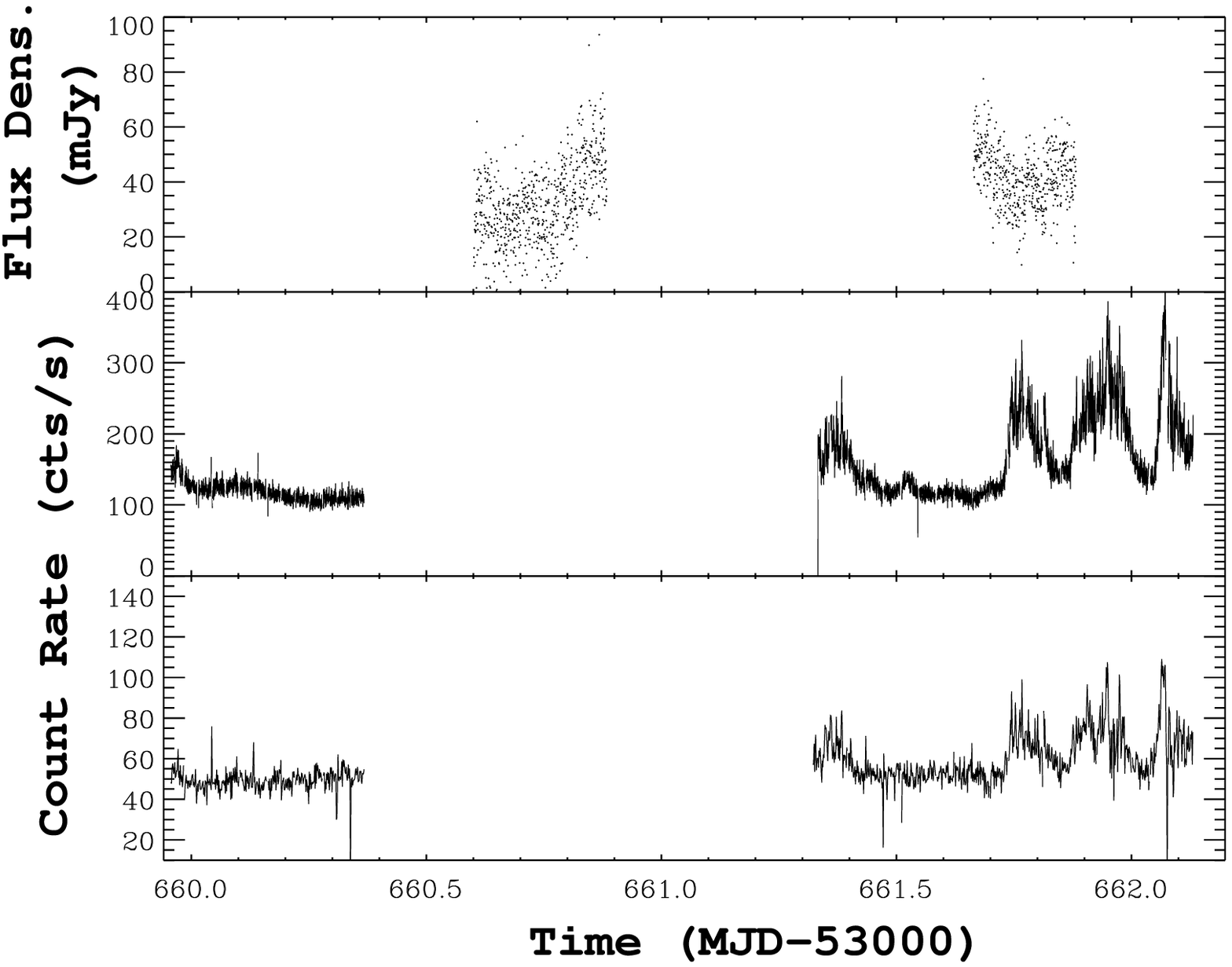}{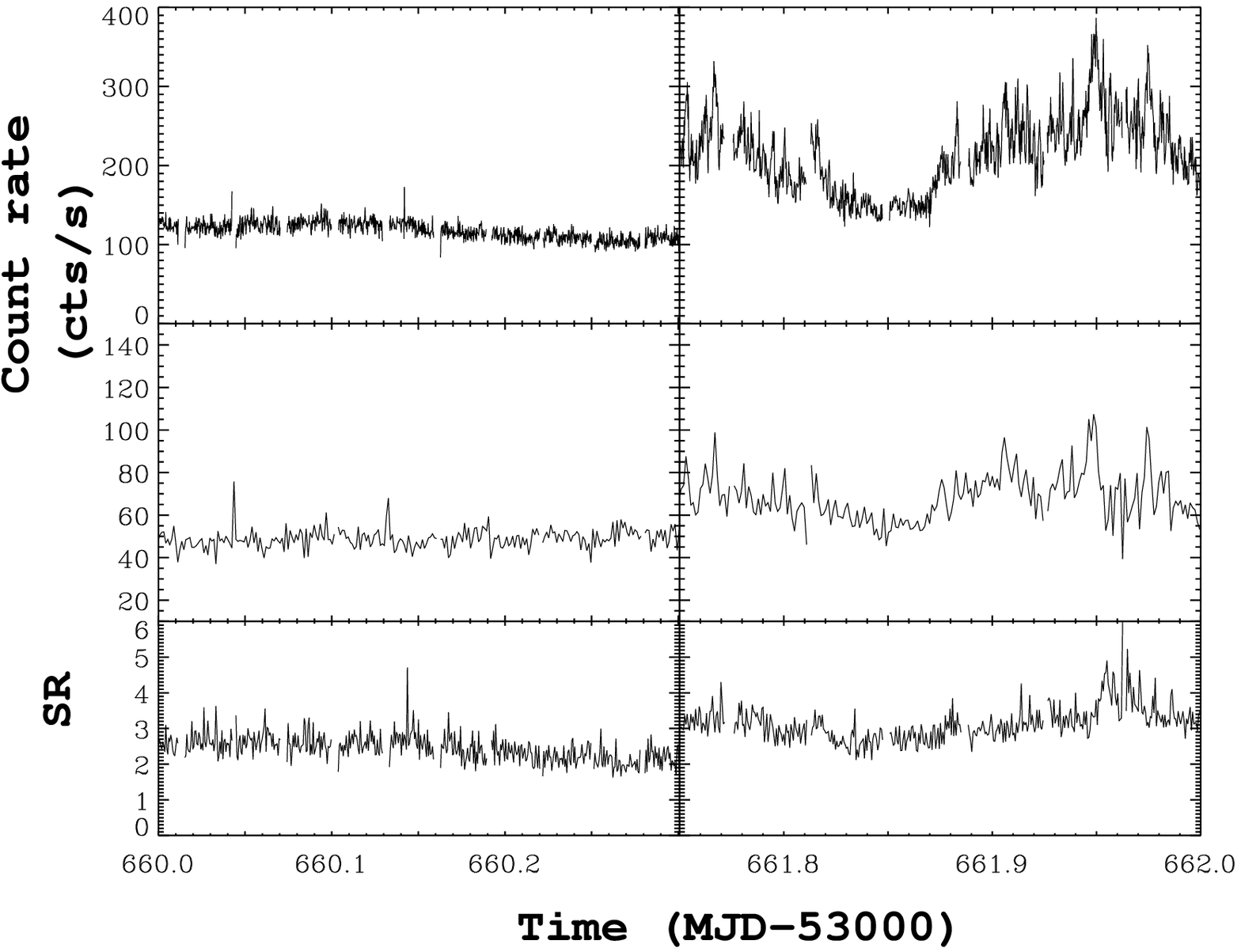}
\label{fig:Oct05}
\end{figure}

\begin{figure}
\figurenum{13}
\caption{{\bf{Left:}} Light curves of \grs\ during Obs. 10: a$)$ RT at 15 GHz, 
b) JEM-X 3--13 keV, c) ISGRI 18--50 keV. {\bf{Right:}} Zoom on 3 sub intervals 
\correc{of Obs. 10 showing (possible) occurences of class $\delta$, $\mu$, and $\beta$. 
From top to bottom the panels respectively show the JEM-X 3--13~keV,
and ISGRI 18--50~keV light curves and the 3--13~keV/18--50~keV SR. }}
\plottwo{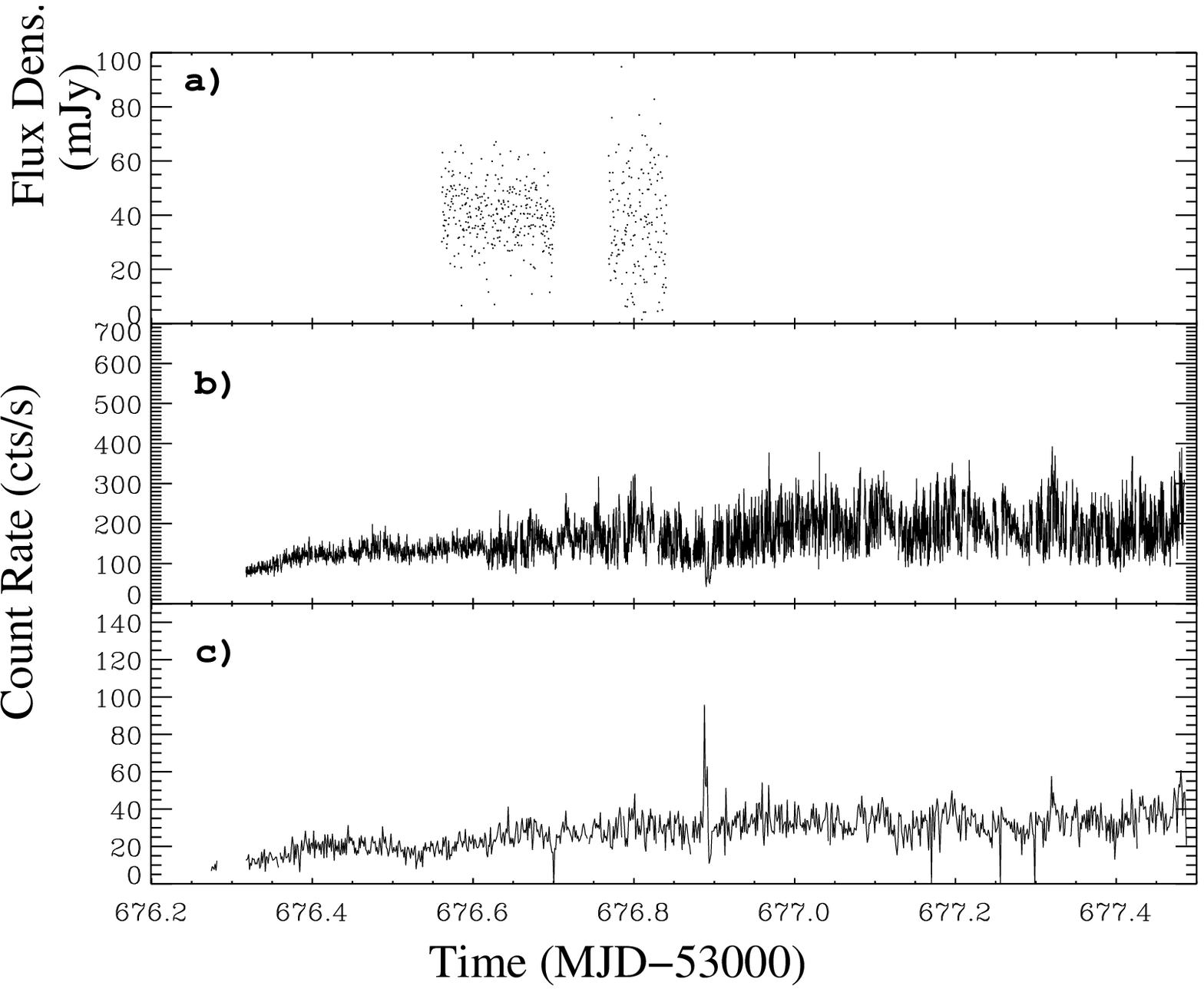}{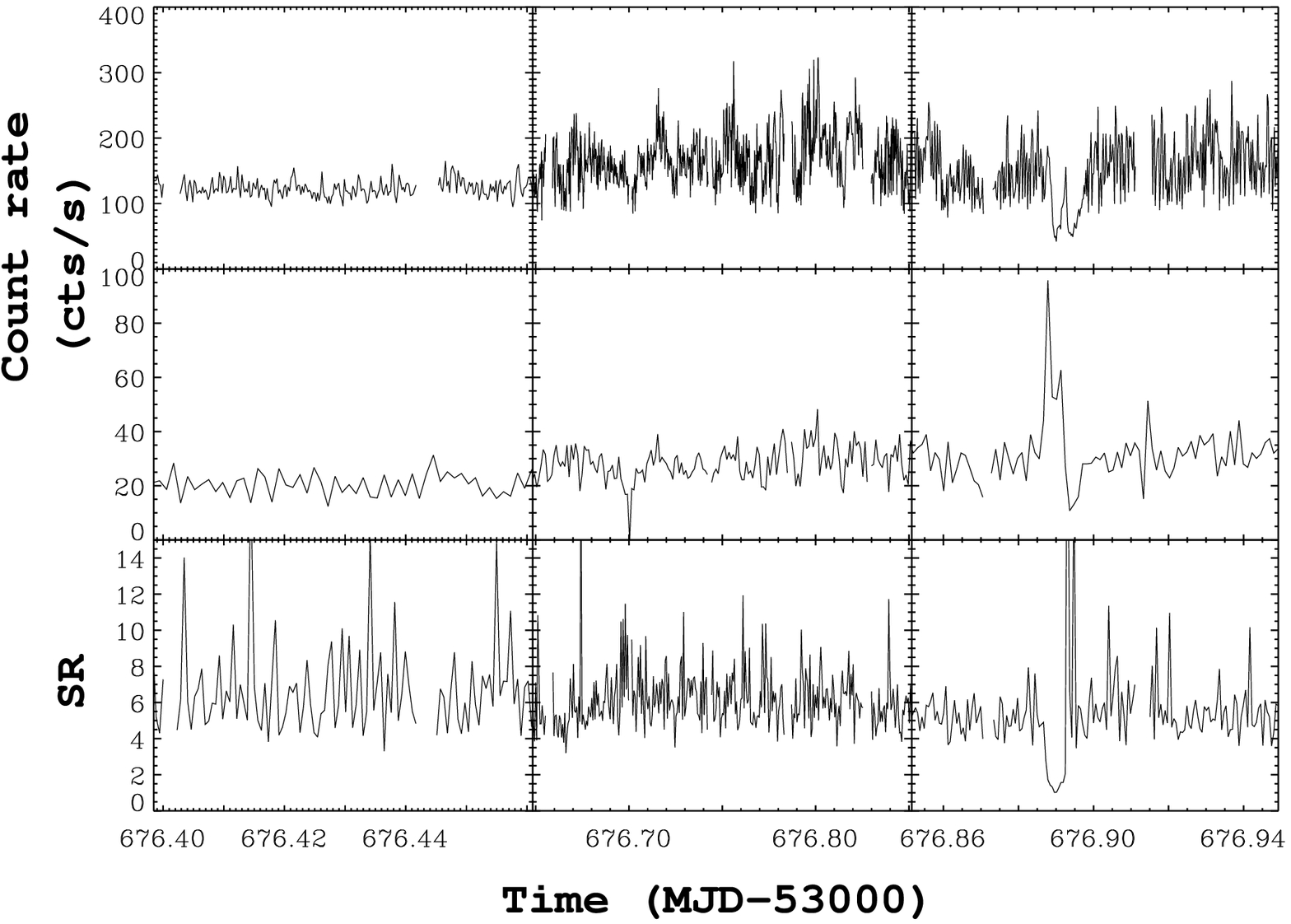}
\label{fig:Nov05}
\end{figure}

\begin{figure}
\figurenum{14}
\caption{{\bf{Left:}} Light curves of \grs\ during Obs. 11: a$)$ JEM-X 3--13 keV \correc{binned at 20~s}, b$)$ ISGRI \correc{18--50 keV binned at 100~s}
 c) PCA \correc{2--18 keV binned at 16~s}. {\bf{Right:}} Zoom on 3 sub intervals showing occurrences of class \correc{$\delta$ (possibly)}, $\mu$ and $\beta$.}
\plottwo{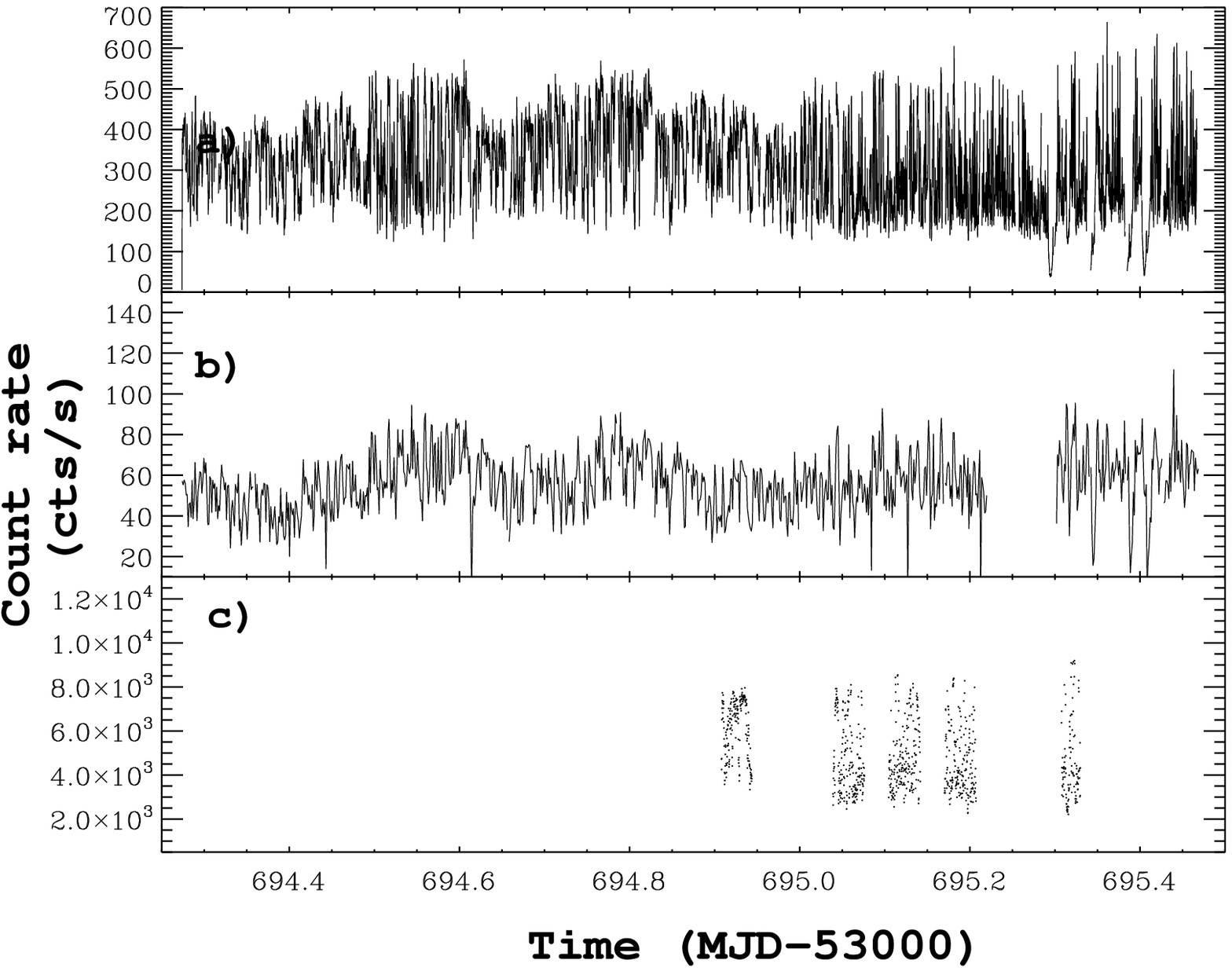}{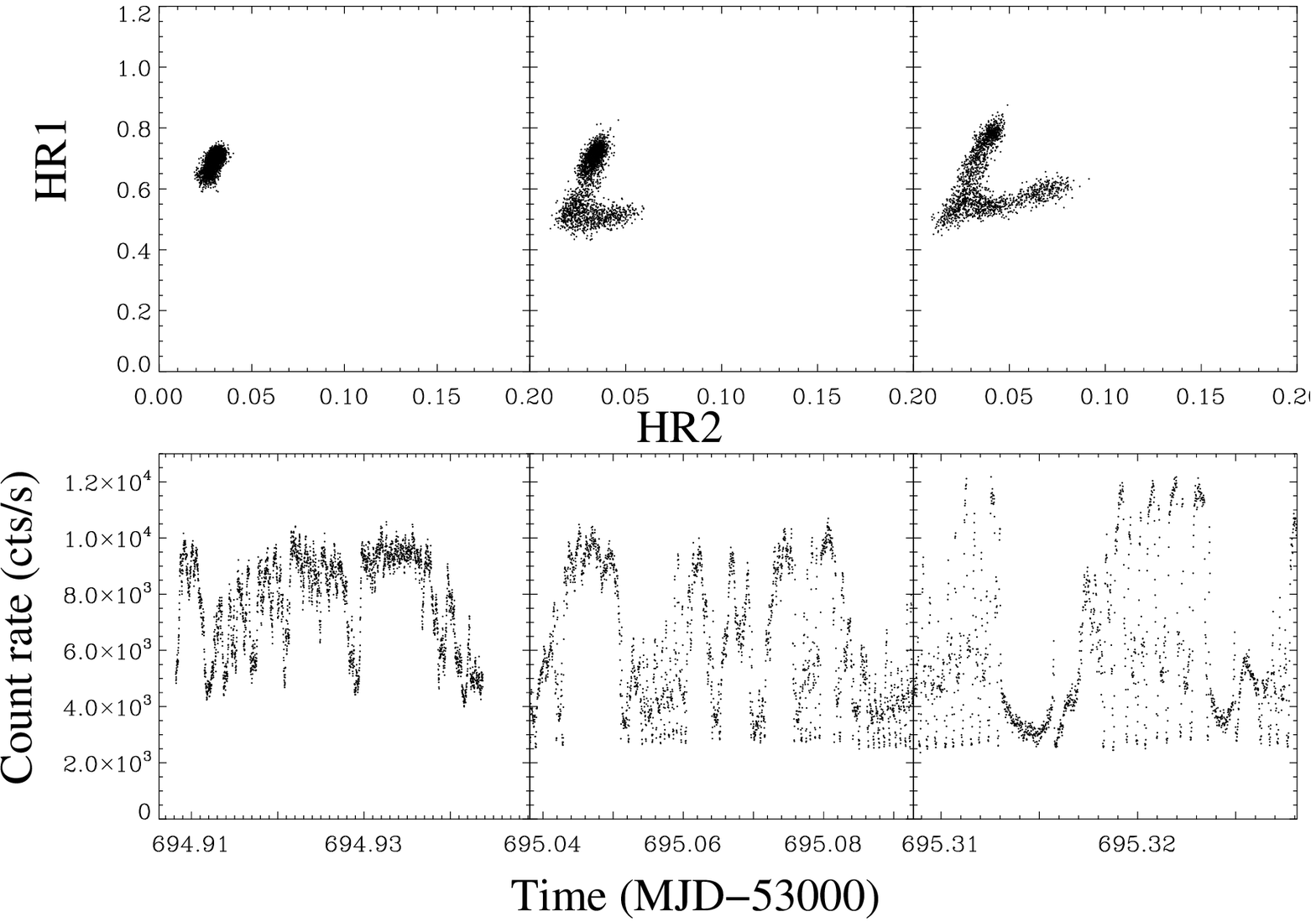}
\label{fig:Nov05_2}
\end{figure}

In all cases the identification of states is based on the simultaneous inspection of
the JEM-X light curves for the general shape, and PCA (when available) for confirmation
through inspection of the CC diagrams. The different responses of JEM-X and PCA 
prevent direct comparison of the CCs produced with the 2 instruments, as they 
lead to completely different patterns. 
In some cases, especially when no simultaneous coverage by \rxte\ is available we identify 
the class as the most likely based on the resemblance with the patterns identified 
by \citet{belloni00}. The multiwavelength light curves and CC diagrams 
are reported in Fig. \ref{fig:Oct04} to \ref{fig:May05} for the four observations
discussed in details in the paper and \ref{fig:Sept05} to \ref{fig:Nov05_2} for the others.
As can be seen, while \rxte\ gives an easy access to high time resolution light curves and 
CC diagrams, \integral\ allows us to have a continuous coverage of the source
and thus study the evolution of classes while witnessing transition between some of them. \\

\indent Obs. 1  (Fig. \ref{fig:Oct04}) shows alternating X-ray dips and spikes (cycles), 
followed by 
intervals of high level of X-ray flux \correc{at all energies}, and variability of 
different duration. Deep inspection 
of these behaviors shows that the source was in an alternation of class $\nu$ and class 
$\rho$-heartbeat, as illustrated in  Fig. \ref{fig:Oct04}. As mentioned by \citet{belloni00},
 each $\nu$-type cycle is separated by the preceding one by an interval of $\rho$ type 
behavior. This 
is illustrated by the fact that the CC diagram (Fig. \ref{fig:Oct04}) of the class 
$\nu$ shows the same pattern as the one from class $\rho$ with an extension towards higher values 
of HR2, corresponding to the dips. The radio light curve shows that 
at least after two sequences of X-ray dip/spike (hereafter cycles) radio flares occured. 
\correc{In the past, similar radio flares have been seen to occur after each cycle 
\citep[e.g.][]{klein02}. This observation is one of the 
four whose deep analysed is presented in the core of this paper.} \\

\indent Obs. 2 (Fig. \ref{fig:Nov04}) shows a high level of very variable X-ray
 emission. On some occasions, one can see short dips in the soft X-ray light curves 
each followed by a spike. \correc{A very variable behavior is also 
seen at hard X-ray energies, with occurrences of short dip clearly visible in the 18--50~keV
light curve (Fig. \ref{fig:Nov04}).} Looking more carefully at the source 
light curves, one can see that \grs\ seems to transit between different 
classes as illustrated by the \rxte\ intervals (Fig. \ref{fig:Nov04}). 
Inspection of the CC diagrams shows that it started in class $\mu$, transited 
to a class $\lambda$, transited back to $\mu$, 
and evolved to a class $\gamma$. The presence of short $\sim100$s long X-ray dips visible in 
the \integral/JEM-X light curves indicate the source transits continuously from $\mu$ to 
$\lambda$. \correc{Deep analysis of this observation is presented in the core of this paper.}\\

\begin{figure}
\figurenum{9}
\epsscale{1}
\caption{{\bf{Left:}} \correc{Light curves of \grs\ during Obs. 3: a$)$  
JEM-X 3--13 keV binned at 50~s, b) 
ISGRI 18--50 keV binned at 200~s, c) \rxte/PCA 2--18~keV binned at 16~s}. {\bf{Right:}}  CC diagram
 (upper panel) and \rxte\ light curve (lower panel) of a sub-interval from Obs. 3 showing 
the source was in class $\chi$.}
\plottwo{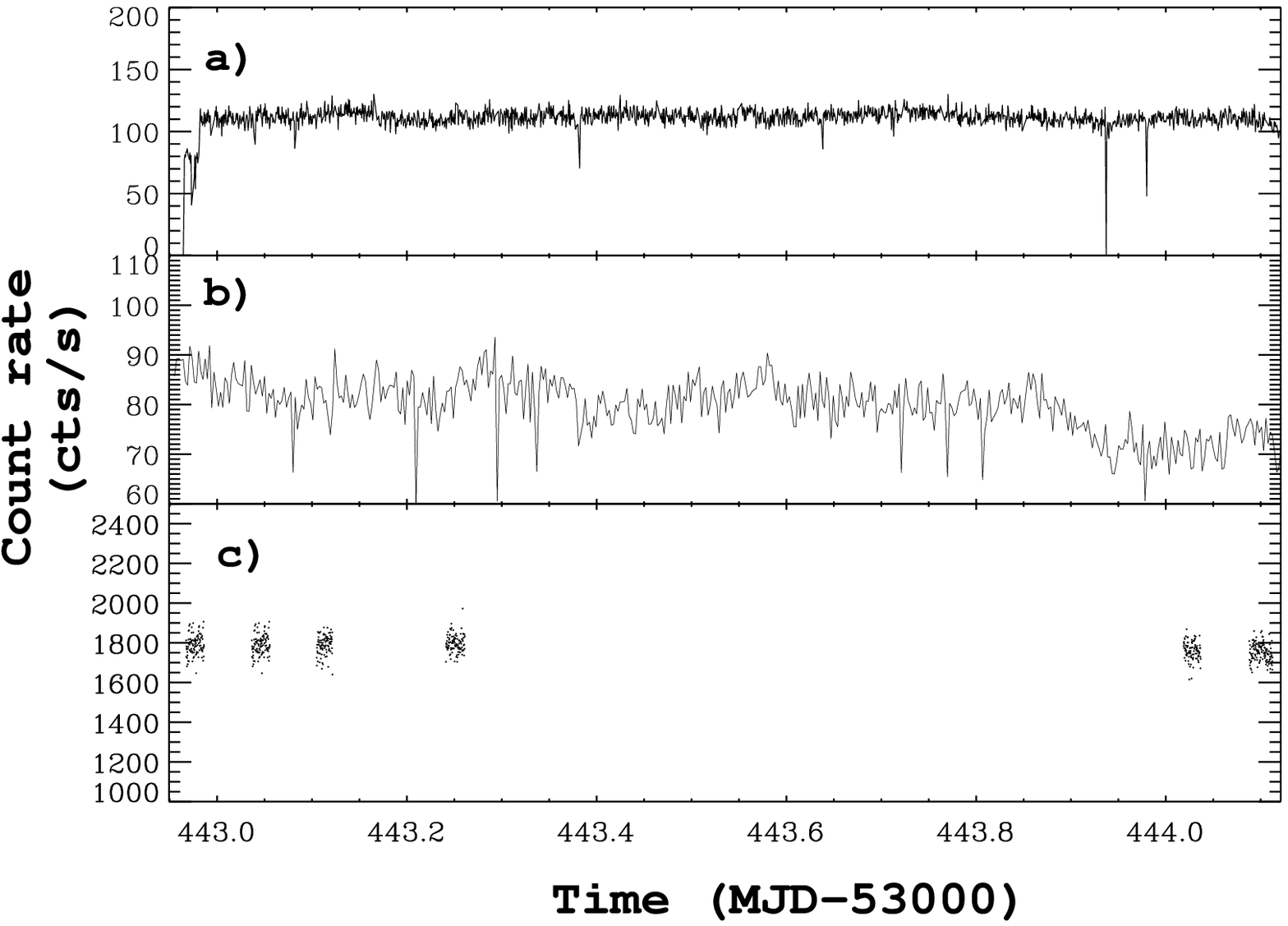}{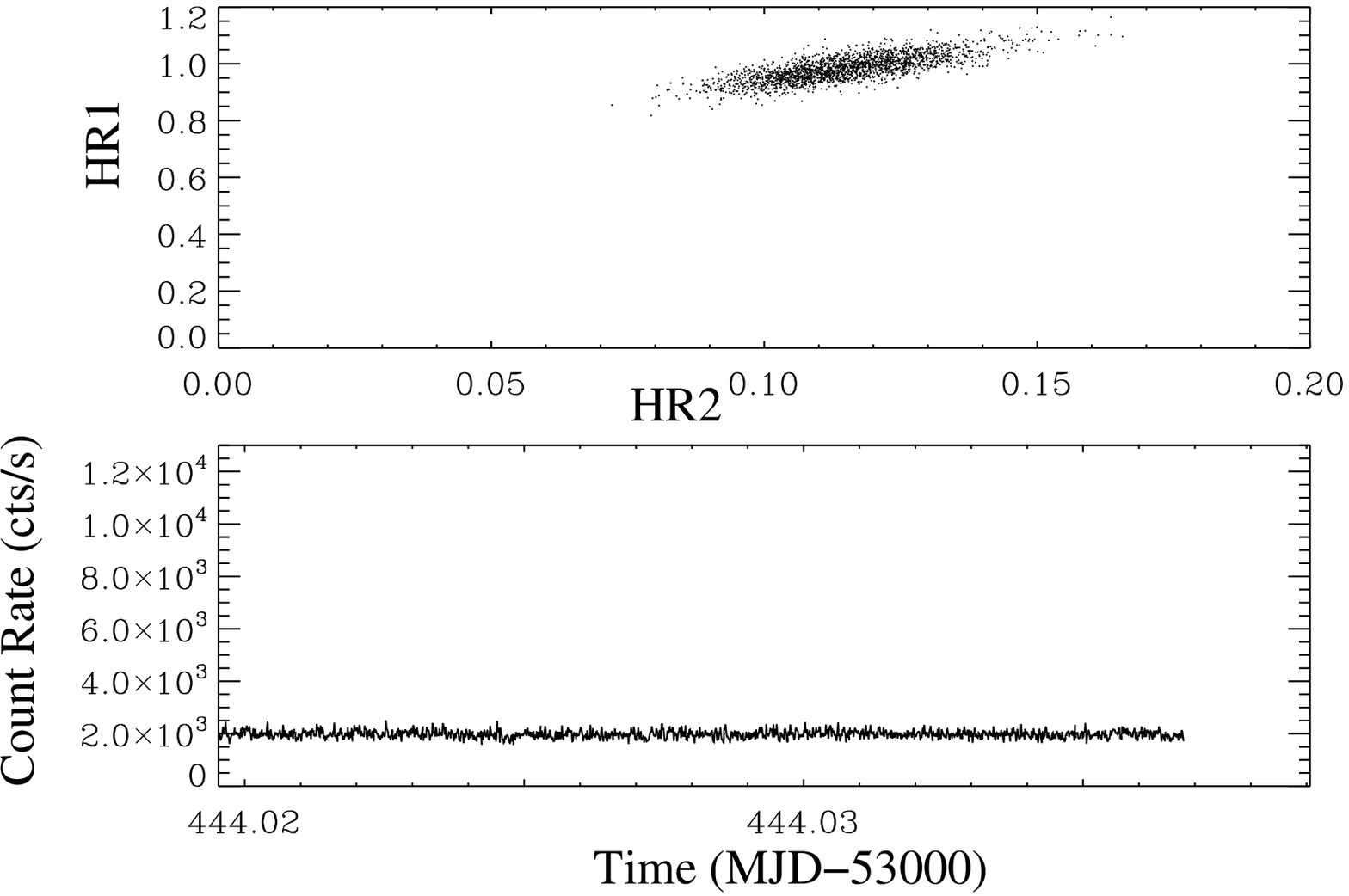}
\label{fig:March04}
\end{figure}

\indent In Obs. 3 (Fig. \ref{fig:March04}) \grs\ 
was in a low luminosity steady state. The 3-13 keV JEM-X 
count rate was around 110 cts/s, \correc{while the 18--50~keV count rate was 
around 80~cts/s, with a slight decrease to about 70~cts/s near the end of the 
observation}. The \correc{2--18}~keV PCA count rate was around 
1800~cts/s. \correc{All light curves show a high degree of rapid variability}. 
Both the light curve and the CC diagrams indicate the source was in class $\chi$. 
No RT were performed during this interval. \correc{The values of HR1 and HR2 are indicative
of a very hard class $\chi$ (Fig. \ref{fig:March04}). In particular, the source is much harder here
than during Obs. 4, although both observations belong to the same class.}\\

\begin{figure}
\epsscale{0.5}
\figurenum{1}
\caption{{\bf{Top:}} Light curves of \grs\ from MJD 53 200 to MJD 53 900 as seen a$)$ between 1.2 and 
12 keV with the \rxte/ASM, and b$)$ at 15 GHz with the RT. In each panel the vertical arrows show the dates of the \integral\ observations, and the longer arrows 
show the four particular observations whose accretion ejection properties are discussed in detail in
 this paper.}
\plotone{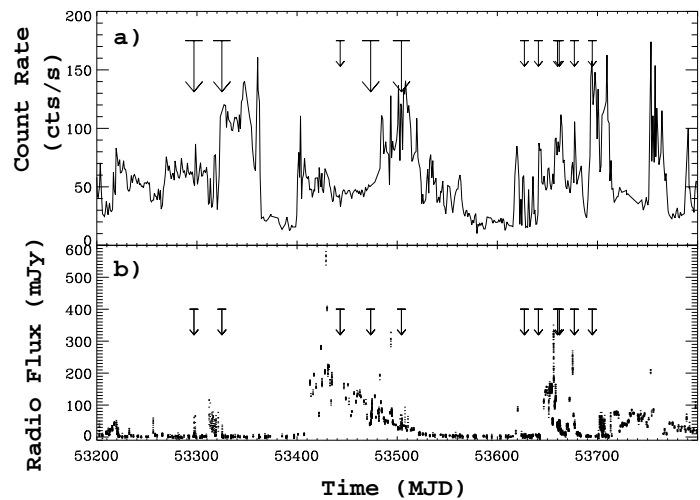}
\label{fig:lite}
\end{figure}

\indent Obs. 4 (Fig. \ref{fig:April04}) is very similar to Obs. 3 although 
\grs\ was  slightly fainter at hard X-ray energies \correc{and showed lower 
values of the  HRs}. Near MJD 53474.1 the source fluxes increased suddenly. This indicates it 
underwent a transition to a brighter class, confirmed by the variations of the ASM light curve
(Fig. \ref{fig:lite}). Simultaneous coverage at radio wavelengths shows a rather steady flux 
at around 50 mJy, although some variations are visible. In particular the radio flux increased
up to 100 mJy near the end of the observation (Fig. \ref{fig:April04}). The global X-ray behavior
indicates that \grs\ was in a radio loud class $\chi$. \correc{This observation is one of the 
four that is deeply analysed in this paper.}\\

\indent Obs. 5 (Fig. \ref{fig:May05}) shows that \grs\ was in classes of high X-ray flux 
 with a high degree of variability. \correc{It is interesting to note, that the overall 
18--50~keV variability increased along the observation. Although the mean 18--50~keV flux 
remained quite bright around 40--50 cts/s, significant dips and spikes can be seen, some 
in simultaneity with those seen at soft X-rays.} It is clear from 
Fig.~\ref{fig:May05} that it transited 
at least between two such classes. Inspection of the CC diagrams shows that it started 
in a class $\mu$ and finished in class $\beta$. Our radio observation shows a roughly 
constant flux of $\sim 40$ mJy up to MJD 53504.35 where two radio flares occured after 
occurrences of dip-spike cycles in the X-ray domain. \correc{This observation is one of the 
four that is deeply analysed in this paper.}\\

\indent Obs. 6 was of poor quality due to the effects of high solar activity and 
was therefore not considered in this analysis.\\

\indent Obs. 7 (Fig. \ref{fig:Sept05}) shows a clear transition from a class with a very
 low level of X-ray emission, to a class of high flux, with high variability \correc{with
similar evolving patterns at both soft and hard X-rays.} The change from 
the first class to the last seems to occur through a third one, as illustrated by the differences 
in the JEM-X \correc{and ISGRI} light curves, but the lack of \rxte\ data during this 
interval prevents us to obtain 
any clear identification. \correc{A zoom on the intermediate part (Fig. \ref{fig:Sep05Theta}) 
shows occurences of 3--13~keV ``M-shape'' patterns typical of class $\theta$. 
In the mean time, the 18--50~keV light curve shows similar variability (the ``M'' can 
be seen in coincidence with those occurring at soft X-rays) and a high 
variability. The SR indicates the source was hard during the dips of the ``M'', which 
is in agreement with the known behavior of  class $\theta$\citep{belloni00}. This class 
shows a behavior very similar to other classes with cycles, i.e. spectrally hard dips ended 
by a soft X-ray spike marking the disappearance of the hard X-ray emission (and transition 
to a soft state as illustrated by the variations of the SR visible in Fig.~\ref{fig:Sep05Theta}).}
 The level of radio emission was quite low, with some variability, \correc{which is also
in agreement with \grs\ being in class $\theta$ \citep{klein02}. This may further corroborates
the generalization of the X-ray to radio connection presented in this paper, although
the resolution at radio wavelength do not allow us to identify any radio flares.} 
 The analysis of the \rxte\ data of the begining and end of the \integral\ observations 
(Fig. \ref{fig:Sept05}) indicates that \grs\ started in class $\phi$ and ended in class 
$\delta$.\\

\indent Obs. 8 \& 9 (Fig. \ref{fig:Oct05}) were part of a large multi-wavelength campaign 
involving at high energy the {\it{Suzaku}} satellite. \correc{The \integral\  light curves
 are first very similar to the light curves of Obs. 4, which 
was indeed a class $\chi$. The higher level of soft X-ray emission, and slightly 
lower level of hard X-ray emission tend to indicate a rather soft class $\chi$. In the second
part, a flare occured at the beginning of the observing interval at both soft and hard X-rays.
\grs\ then returned to a rather steady emission, before transiting to a class that showed 
similar patterns at soft and hard X-rays. The lack of simultaneous \rxte\ coverage prevents 
to securely identify the class of varibility.}  \correc{A zoom on the JEM-X light curve 
(Fig. \ref{fig:Oct05}) does not show any easily identifiable pattern. Furthermore, the 
3--13~keV/18--50~keV SR has a value quite similar to that of Obs. 8 (Fig. \ref{fig:Oct05}) 
and even to Obs. 4 (not shown). Therefore, it seems the source was still in a
class $\chi$.} \citet{ueda06} presented preliminar results of the whole campaign, 
and they, in particular, showed that \grs\ started in a class $\chi$ 
showing the presence of a 6 Hz QPO, and transited in a class $\theta$. \correc{The class $\theta$,
however, occured between both \integral\ intervals.}   
\correc{The mean level of radio
 emission was 31~mJy until MJD 53660.7 then increased to 
about 39~mJY during the first radio interval (Fig. \ref{fig:Oct05}). The mean level of 
radio emission during the second interval was about 40~mJy. The shape of the radio lightcurve 
may suggest that ejection had taken place between the two observing intervals. A simultaneous
VLA light curve \citep{ueda06} is also compatible with this interpretation.}
\citet{ueda06} suggested that ejection of material was triggered at 
the transition from spectral state-C to spectral state-A (hard intermediate to soft state),
similar to our interpretation. \\

\indent Obs. 10 (Fig. \ref{fig:Nov05}), shows a complex behavior made of several 
possible transitions.  The radio light curve 
(Fig. \ref{fig:Nov05}) also indicates 2 different behaviors. In the first part 
(\correc{MJD~53676.56 to 53676.99}), the \correc{mean 15 GHz
flux was $\sim 40.4$ mJy with an RMS of 10.4~mJy }. In the second part \correc{(MJD~53676.77
 to 53676.841), the mean flux was 36.5~mJy, and the} variability of the radio 
emission had increased significantly \correc{with an RMS of 19.6~mJy. This 
increase seems to be correlated to the X-ray behavior. Around MJD~51678.9 a dip at soft 
X-ray energies associated to a 18--50~keV strong flare occured}. The identification of the 
classes is only based on the shape of the light curves \correc{and 3--13~keV/18--50~keV SR}, 
since no \rxte\ data are available. We can distinguish 3 main periods in the light curves. 
Until MJD 53676.6, the source flux gradually increased and so did the variability. 
The shape of the JEM-X light curve (Fig. \ref{fig:Nov05}, right panel) and the level 
of radio emission seem to indicate a class $\chi$ observation. \correc{The low 
level of hard X-ray emission and the SR, however, do not favor this identification.
Fig. \ref{fig:Nov05} (right) shows that \grs\ was transiting between state 
with SR$\sim4$ and SR$\sim14$--$15$. 
This behavior is not what is observed in the other class $\chi$ observations. This would 
rather indicate a class with transitions between spectral state-A and state-B of \citet{belloni00},
possibly class $\delta$, although it was then fainter than the other 
class $\delta$ observations (e.g. Obs. 5) at both soft and hard X-rays. Note that 
the possible level of radio emission was not what is usually observed in this 
class \citet{klein02} which weakens the identification}. After MJD 53676.6, the shape resembles 
that of a class $\mu$ (Fig. \ref{fig:Nov05}, right panel), before undergoing a 
dip-spike sequence indicative of a class $\beta$. \correc{The identification of class $\mu$
is compatible with the level of  radio emission \citep{klein02}.}
It has to be noted that after the occurrence of this cycle, the level of X-ray emission 
was higher, and more variable than before. \correc{Although there were no more occurrences of 
cycles towards the end of the observation, the X-ray behavior was similar} to what is observed
 during class $\beta$ between the occurrences of cycles  \citep{belloni00} with a high
degree of variability. \\

\indent Obs. 11 (Fig.  \ref{fig:Nov05_2}) also shows several transitions. We do not have any 
radio coverage during this observation. Although our \rxte\ observations arrive late 
in the \integral\ coverage (Fig.  \ref{fig:Nov05_2}), it seems that during the first part 
 of the observation (up to about MJD 53695), \grs\ was in the same type of variability.
Zooms on the JEM-X \correc{and ISGRI} light curves during this first part (not shown) indeed 
indicate similar morphologies. Inspection of the CC diagram of the first sample of the \rxte\ light curve
shows the source was in a class $\delta$ (Fig.~\ref{fig:Nov05_2}, right panel). The
 following samples show that \grs\ had changed classe. This is especially indicated 
by the track in the CC diagram that shows an incursion in the low HR1 region, with an 
HR2 as high as 0.05. This pattern \correc{and the shape of the light curve} is what is 
observed during class  $\mu$. However typical class $\mu$, as the one 
observed during Obs. 3, show a longer extension towards higher value of HR1 (Fig.  
\ref{fig:Nov04} right
panel). This may simply indicate that while in the first hundred seconds of the interval
 the variations were still similar to $\delta$, \grs\ evolved towards $\beta$ in the end, and hence the CC 
diagram is a 
mixture of both, or simply that due to the short length of the interval ($\sim$2500~s) 
the high values 
of HR2 are less observed than in typical ($>$3000s long) intervals. After that, around 
MJD 53695.1, another transition occured.\grs\ shows light curves with a high level of 
X-ray emission and a high degree of variability. \correc{The behavior was the same at soft
and hard X-rays with occurrences of long dips in both light curves.} 
This is reminiscent of what we saw during Obs. 5 (Fig.  \ref{fig:May05}), 
with the occurrence of $\beta$-like cycles after some times 
(Fig.  \ref{fig:Nov05_2}, right panel).

\newpage 
\begin{table}
\tablenum{1}
\caption{Journal of all the \integral\ observations of our campaign. \correc{The simultaneous 
RT observations are also indicated.}}
\begin{tabular}{cccccccc}
\hline
\hline
Obs. \# & Revolution \# & ObsId & \multicolumn{2}{c}{MJD} & \multicolumn{2}{c}{RT Radio observations}\\
      &     &          & Start & Stop & Start & Stop \\
\hline
1 &  246 & 02200280001 & 53296.3730 & 53297.5848 & 53296.710 & 53296.879\\
2 &  255 & 02200280002 & 53324.2773 & 53325.5107 & 53324.519 & 53324.803\\
3 &  295 & 02200280003 & 53442.9542 & 53444.1497 &  \multicolumn{2}{c}{No coverage}\\
4 &  305 & 02200280004 & 53472.8673 & 53474.1419 & 53473.098 & 53473.417\\
\nodata & \nodata & \nodata  & \nodata    & \nodata    & 53474.085 & 53474.141 \\
5 &  315 & 02200280005 & 53503.6328 & 53504.8691 & 53503.972 & 53504.353 \\
6 &  356 & 03200200001 & 53626.3852 & 53627.5983 & 53626.853 & 53626.996\\
7 &  361 & 03200200002 & 53640.3856 & 53641.5803 & 53640.870 & 53640.946\\
8 &  367 & 03200200003 & 53659.9563 & 53660.3679 & \multicolumn{2}{c}{No coverage}\\
9 &  368 & 03200200004 & 53661.3222 & 53662.1313 & 53661.666 & 53661.879\\
10 & 373 & 03200200005 & 53676.2472 & 53677.4881& 53676.562 & 53676.840 \\
11 & 379 & 03200200006 & 53694.2739 & 53695.4691& \multicolumn{2}{c}{No coverage} \\
\hline
\end{tabular}
\label{tab:journalinteg}
\end{table}

\newpage
\begin{table}
\tablenum{2}
\caption{Journal of the \rxte\ observations of our campaign.}
\begin{tabular}{cccc}
\hline
\hline
Obs. \# & ObsId & MJD start & Date\\
(\integral\ equivalent)&  &   (MJD)   & (yyyy-mm-dd)\\
\hline
1 & 90105-03-02-00 & 53296.387 & 2004-10-18\\
1 & 90105-03-02-01 & 53296.593& 2004-10-18\\
1 & 90105-03-02-02 & 53296.661& 2004-10-18\\
1 & 90105-03-02-03 & 53296.728& 2004-10-18\\
1 & 90105-03-02-04 & 53296.794& 2004-10-18\\
1 & 90105-03-01-000 &53297.039 & 2004-10-19\\
1 & 90105-03-01-00 & 53297.370 & 2004-10-19\\
1 & 90105-03-03-00 & 53297.440 & 2004-10-19\\
1 & 90105-03-03-01 & 53297.508 & 2004-10-19\\
1 & 90105-03-03-02 & 53297.576 & 2004-10-19\\
2 & 90105-05-01-00 & 53324.261 & 2004-11-15\\
2 & 90105-05-02-00 & 53324.524 & 2004-11-15\\
2 & 90105-06-01-00 & 53325.180 & 2004-11-16\\
2 & 90105-06-02-00 & 53325.442 & 2004-11-16\\
3 & 90105-05-03-00 & 53442.968 & 2005-03-13\\
3 & 90105-05-03-01 & 53443.037 & 2005-03-14\\
3 & 90105-05-03-02 & 53443.105 & 2005-03-14\\
3 & 90105-05-03-03 & 53443.242 & 2005-03-14\\
3 & 90105-05-03-04 & 53444.019 & 2005-03-15\\
3 & 90105-05-03-05 & 53444.089 & 2005-03-15\\
4 & 90105-07-01-00 & 53472.921 & 2005-04-12\\
4 & 90105-07-02-00 & 53473.054 & 2005-04-13\\
4 & 90105-07-03-00 & 53473.972 & 2005-04-13\\
5 & 90105-08-01-00 & 53503.669 & 2005-05-13\\
5 & 90105-08-02-00 & 53503.870 & 2005-05-13\\
5 & 90105-08-03-00 & 53504.719 & 2005-05-14\\
7 & 90105-04-01-00 & 53640.390 & 2005-09-27\\
7 & 90105-04-02-00 & 53641.111 & 2005-09-28\\
7 & 90105-04-03-00 & 53641.439 & 2005-09-28\\
7 & 90105-04-03-01 & 53641.516 & 2005-09-28\\
11 & 90105-06-03-01 & 53694.908 & 2005-11-20\\
11 & 90105-06-03-00 & 53695.039 & 2005-11-21\\
11 & 90105-06-03-02 & 53695.308 & 2005-11-21\\

\hline
\end{tabular}
\label{tab:logrxte}
\end{table}

\newpage
\begin{table}
\tablenum{3}
\caption{\correc{``Quick-look'' results of our campaign:} list of sources detected by ISGRI \correc (with a 
signal to noise ratio greater than 6) \correc{and classes of variability of \grs\ in each observation.} \label{tab:Listsource}}
\begin{tabular}{l}
\epsfig{file=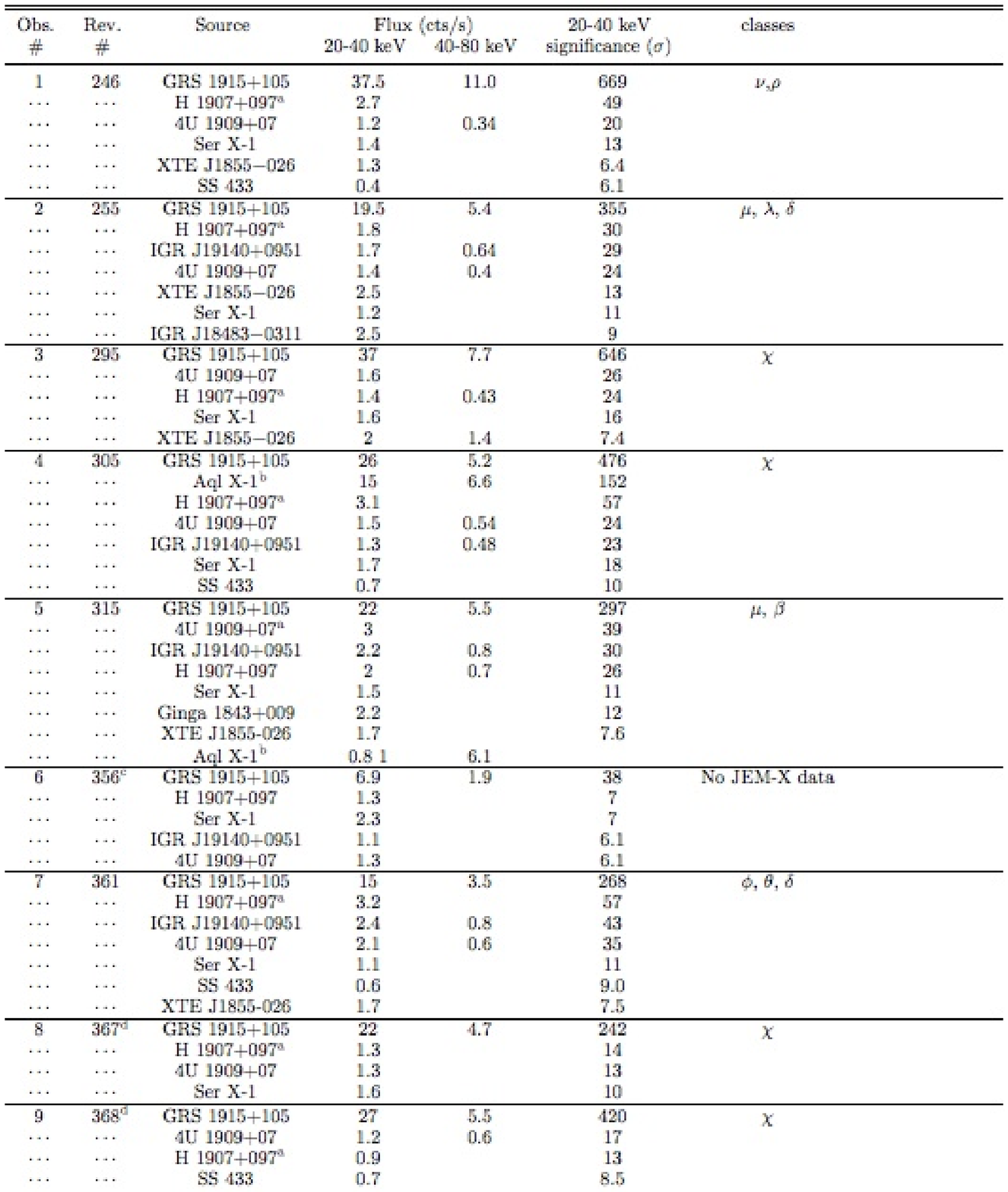,width=19cm}\\
\end{tabular}
\end{table}

\begin{table}
\begin{tabular}{l}
\epsfig{file=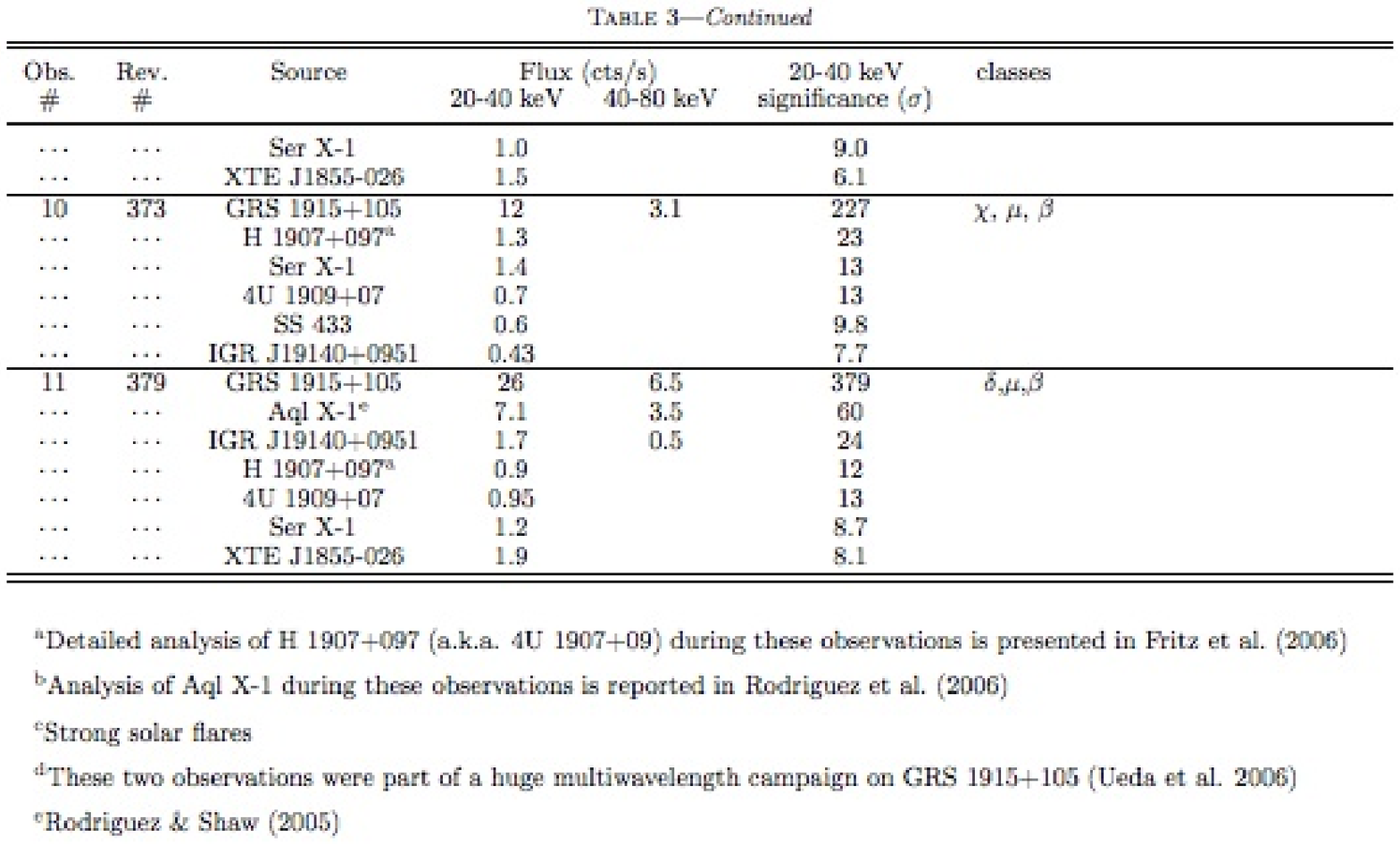,width=19cm}\\
\end{tabular}
\end{table}

\newpage

\newpage

\newpage

\newpage

\newpage

\newpage

\newpage

\newpage

\newpage

\newpage

\newpage

\end{document}